\newif \iftwocol
\twocolfalse
\iftwocol
\documentclass[11pt,twocolumn]{IEEEtran}
\else
\documentclass[letter,10pt,onecolumn]{IEEEtran}
\fi

\usepackage{psfrag}

\usepackage{amsmath}
\usepackage{amsthm}
\usepackage{amssymb}
\usepackage{cite}
\usepackage{graphicx}
\usepackage{booktabs}
\usepackage{subfigure}
\usepackage{color}
\usepackage{url}

\newcommand{\BIT}{\begin{itemize}}
\newcommand{\EIT}{\end{itemize}}
\newcommand{\BEN}{\begin{enumerate}}
\newcommand{\EEN}{\end{enumerate}}

\newcommand{\N}{\mathcal{N}}

\newcommand{\Li}{\mathcal{L}}

\newcommand{\B}{\mathcal{B}}
\newcommand{\A}{\mathcal{A}}

\newcommand{\yb}{\mathbf{y}}

\newcommand{\xb}{\mathbf{x}}

\newcommand{\Ex}{\mathbb{E}}
\providecommand{\norm}[1]{\left\lvert#1\right\rvert^2}
\providecommand{\abs}[1]{\left\lvert#1\right\rvert}

\newcommand{\ra}{\rightarrow}

\newcommand{\NoN}{\nonumber}

\newcommand{\INR}{{\textsf{INR}}}
\newcommand{\SNR}{{\textsf{SNR}}}

\DeclareMathOperator{\var}{var}
\newcommand{\EPS}{{A_{\epsilon}^*}^n}

\newtheorem{remark}{Remark}
\newtheorem{theorem}{Theorem}
\newtheorem{corollary}{Corollary}

\begin{document}
\title{Two Birds and One Stone: Gaussian Interference Channel with a Shared Out-of-Band Relay of Limited Rate}
\author{{Peyman Razaghi, Song Nam Hong,  Lei Zhou, Wei Yu, and Giuseppe Caire}
\thanks{Manuscript submitted on April 3, 2011, revised on Sep 1, 2012. Part of this paper was presented in 2010 Information Theory and Applications Workshop, University of California San Diego, San Diego \cite{peyman_yu_ita10}. Peyman Razaghi (razaghi@usc.edu), Song Nam Hong (songnamh@usc.edu) and
Giuseppe Caire (caire@usc.edu) are with the University of
Southern California, and Lei Zhou (zhoulei@comm.utoronto.ca) and Wei Yu
(weiyu@comm.utoronto.ca) are with the University of Toronto. Kindly please
address correspondence to Giuseppe Caire at caire@usc.edu.}
}
\maketitle

\begin{abstract}
The two-user Gaussian interference channel with a shared out-of-band relay is considered. The relay observes a linear combination of the source signals and
broadcasts a common message to the two destinations, through a perfect link of fixed limited rate $R_0$ bits per channel use. The out-of-band nature of the relay is reflected by the fact that the common relay message does not interfere with the received signal at the two destinations. A general achievable  rate is established,  along with upper bounds on the capacity region for the Gaussian case.  For $R_0$ values below a certain threshold, which depends on channel parameters, the capacity region of this channel is determined in this paper to within a constant gap of $\Delta=1.95$ bits.  We identify interference regimes where a two-for-one gain in achievable rates is possible for every bit relayed,  up to a constant approximation error.  Instrumental to these results is  a carefully-designed quantize-and-forward type of relay strategy along with a joint decoding scheme employed at destination ends. Further, we also study successive decoding strategies with optimal decoding order (corresponding to the order at which common, private, and relay messages are decoded), and show that successive decoding also achieves  two-for-one  gains asymptotically in regimes where a two-for-one gain is achievable by joint decoding; yet,  successive decoding produces unbounded loss asymptotically when compared to joint decoding, in general.




\end{abstract}


\section{Introduction}
The butterfly network \cite{ahlswede_network_coding}, the coat of arms of network coding, exemplifies a fascinating fact about networks: A single relayed bit may turn into multiple information bits at different destination. In other words, the same relayed message conveys different information in different side-information contexts. Yet, there are quite many restrictions to have such efficiency in digital network coding. First, the two-for-one gain in the butterfly network example holds in a  ``multi-source multicast'' scenario, i.e., all destinations decode the message of all sources \cite{li_li_network_coding}.  Then, there is no  noise, and more importantly, there is no interaction between links, for example in the form of interference.

\begin{figure}
\centering
\includegraphics[width=0.7\textwidth]{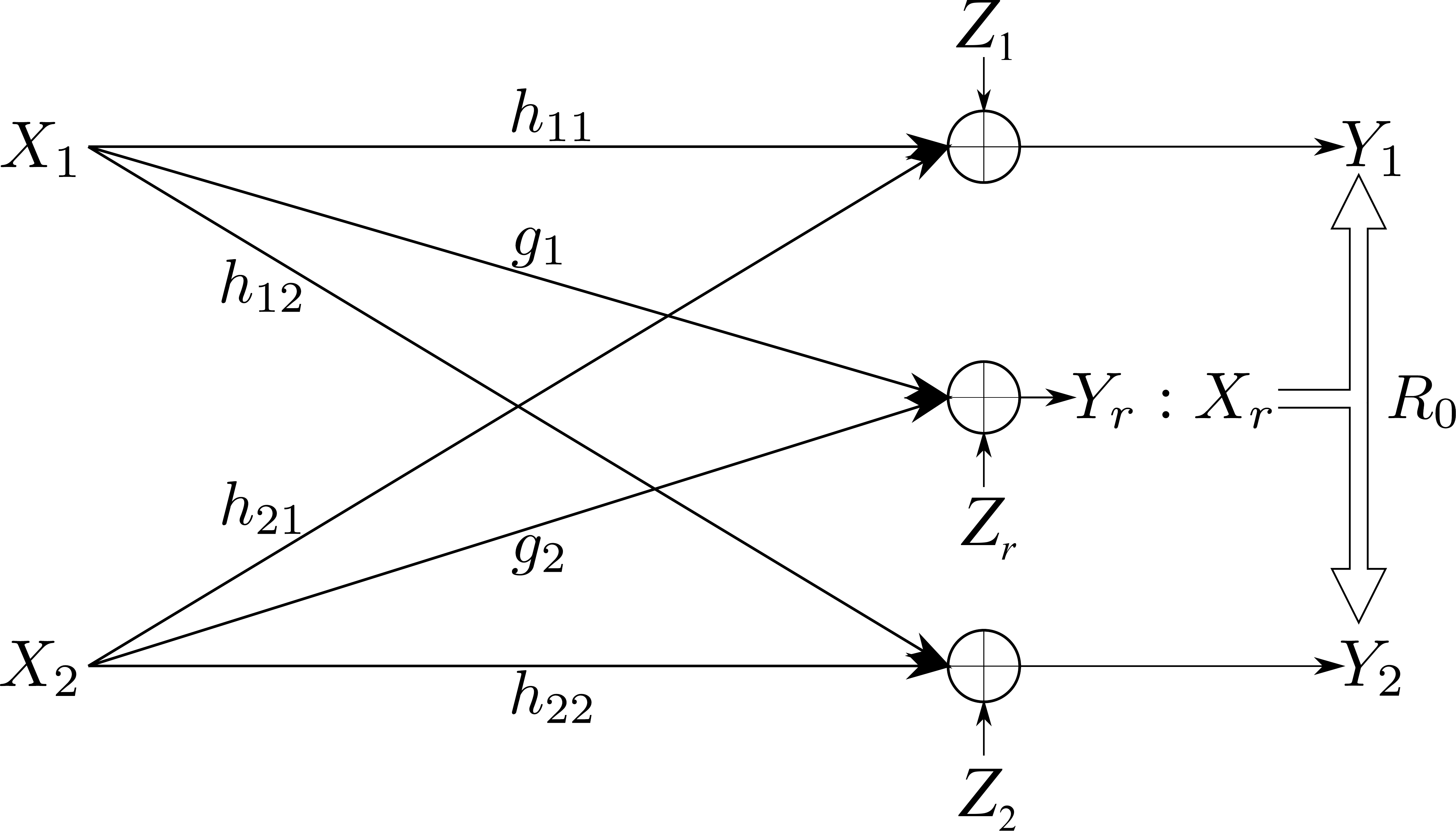}
\caption{A Gaussian interference channel with an out-of-band relay of rate $R_0$. The relay broadcasts a common message to both destinations. \label{fig:channel-model}}
\end{figure}

In wireless channels, a shared relay helping  two destination nodes with a common message resembles a scenario parallel to the butterfly network in network coding,  with noise and interference representing subtle differentiating factors. In the wireless case, an equivalent problem is to find relay strategies that simultaneously assist both destinations.  Ideally, we would like that for every bit relayed, the achievable rate to each destination improves by one bit. However, as one might expect, such two-for-one improvements may not be always achievable in the wireless scenario, particularly due to presence of interference and noise.

Consider a multi-source unicast scenario represented by a two-user Gaussian interference relay channel augmented with an out-of-band relay as shown in Fig.~\ref{fig:channel-model}.  
The channel is defined as:
\begin{subequations}
  \label{eq:7}
 \begin{align}
  Y_1&=h_{11}X_1+h_{21}X_2+Z_1\\
  Y_2&=h_{12}X_1+h_{22}X_2+Z_2\\
\intertext{and}
  Y_r&=g_1 X_1+g_2X_2+Z_r,
\end{align}
\end{subequations}
where $X_1$ and $X_2$ are the transmitted symbols with powers
$P_1 = E[|X_1|^2]$ and $P_2 = E[|X_2|^2]$, $Y_1$ and $Y_2$ and $Y_r$ denote the channel outputs at destinations 1 and 2 and at the relay, respectively, and $Z_1$, $Z_2$ and $Z_r$ denote the corresponding additive-white-Gaussian-noise (AWGN) samples, assumed i.i.d.$\sim\N(0,N)$.  Formally, for given block length $n$, user $i$, $i=1,2$, communicates a random message $m_i$ taken from $\{1,\ldots,2^{nR_i}\}$ by transmitting a codeword $\xb_i(m_i)$ is length $n$ from codebook $\mathcal{C}_i$ of size $2^{nR_i}$, satisfying an average power constraint of $P_i$ such that $\sum_k|\xb_i^k(m_i)|^2/n\leq P_i$, where $k$ is the time index.   The relay observes a sequence of channel outputs $\yb_r$, and in time $k$, transmits a digital message $x_r^k\in \{1,2,\ldots,2^{R_0}\}$, i.e., at rate $R_0$ bits per channel use. The relay message $x_r^k$ is a causal function of past channel outputs at the relay, i.e., $x_r^k = f_r(y_r^1,\ldots,y_r^{k-1})$ for a function $f_r(\cdot)$.  Destination $i$ decodes $\hat{m}_i$ as the transmitted source message based on channel outputs $\yb_i$ and received relay message sequence $\xb_r$, using a decoding function $f_{i,n}: \mathbb{R}^n\times \left\{1,2,\ldots,2^{R_0}\right\}^n\ra \mathbb{N}$. An error occurs if $\hat{m}_i\neq m_i$.  A rate pair $(R_1,R_2)$ is achievable if there exists a sequence of codebooks $\mathcal{C}_1,\mathcal{C}_2$ satisfying the power constraints, and a pair of decoding functions $f_{1,n}(\cdot),f_{2,n}(\cdot)$, such that the expected decoding error probabilities taken with respect to random choice of the transmitted pair of messages tend to zero as $n$ tends to infinity.

\begin{figure}[t]
  \begin{center}
   \centering
   \includegraphics[width=0.7\textwidth]{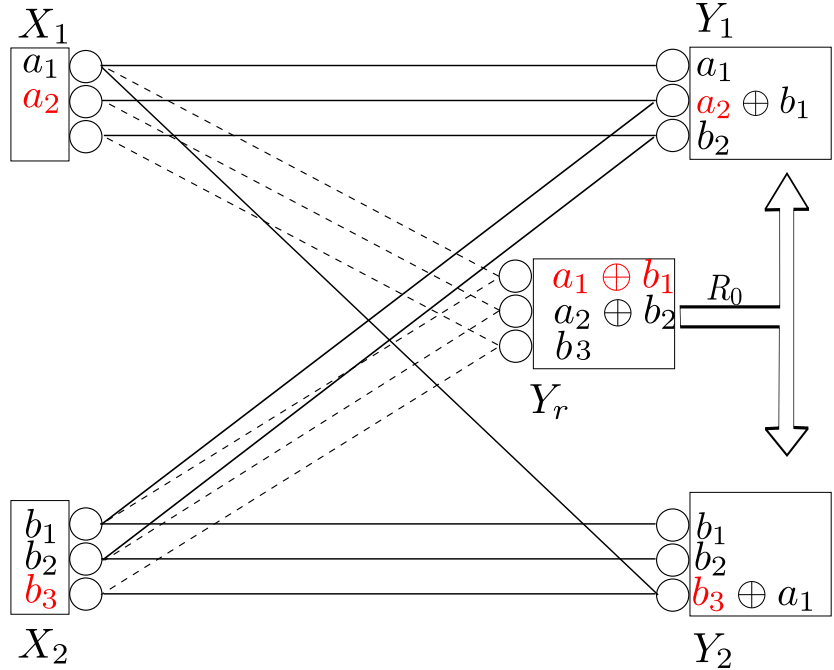}
  \end{center}
  \caption{With the relay forwarding one bit of information to both destinations, each user can send an additional bit.  In the above example, relay forwards $a_1\oplus b_1$, enabling user one to send  $a_2$  and  user two to send $b_3$. \label{fig:det-relay-1}
}
 \end{figure}
\begin{figure}[t]
  \begin{center}
   \centering
   \includegraphics[width=0.7\textwidth]{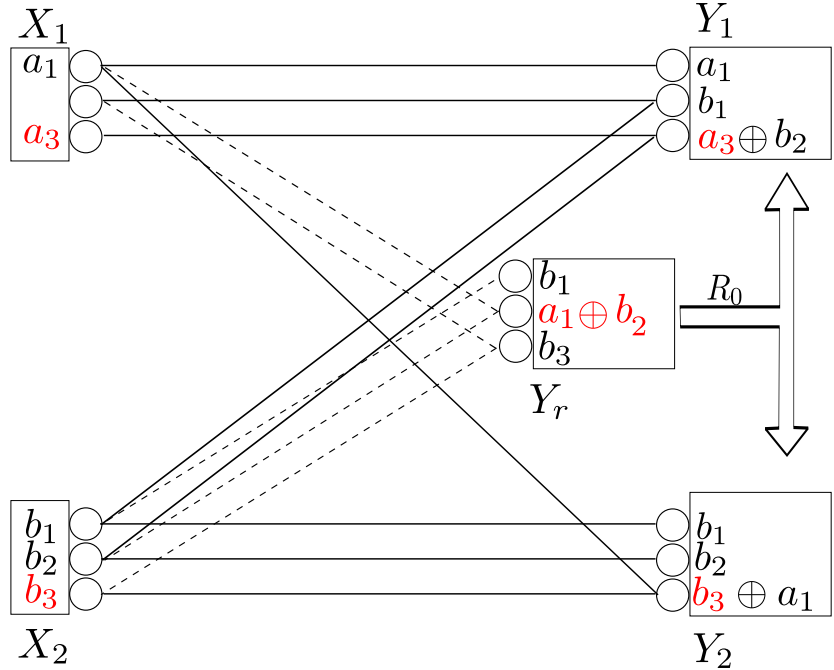}
  \end{center}
  \caption{Relay's quantization strategy depends substantially on channel configuration.  In the above example, the relay forwards its second bit-level $a_1\oplus b_2$, enabling user one to send additional bit $a_3$, and  user two to send additional bit $b_3$. \label{fig:det-relay-2}
}
 \end{figure}

How could the relay assist both users simultaneously? Following along conventional decode-and-forward \cite[Theorem~1]{cover_elgamal} and compress-and-forward relay strategies \cite[Theorem~6]{cover_elgamal}, the relay could decode one or both of the source messages,  or attempt to share a compressed version of its observation with destinations.  Both kind of these strategies are viable (see for example \cite{Maric_Dabora,mdg_2009,sahin_simeone_erkip,ty_2011,ty_2012}), with some limitations. Decode-and-forward type of strategies suffer, for example, when the two messages interfere  more strongly at the relay than at destinations.  Even when the relay has a strong channel to decode one  of the source messages, forwarding a relay message containing information about only one user's message  may not be optimal simultaneously for both users. For example, when interference is very weak,  information about interfering message is of limited value, or when interference is strong, the user can decode and cancel interfering signal with little relay help; see also \cite{Maric_Dabora}, and the two examples in Fig.~\ref{fig:det-relay-1} and Fig.~\ref{fig:det-relay-2}. On the other hand, channel strength disparities are problematic for compress-and-forward type of strategies where the  relay communicates its compressed observation to both destinations using a single message. From the relay's perspective of compressing its received signal, an asymmetric channel means one destination can obtain a finer quantized version of the relay observation. As a result, one cannot design a single quantization scheme to faithfully communicate the relay's received signal to the destinations simultaneously; see also \cite{peyman_yu_ita10}.

Yet under certain conditions, it is possible to obtain similar two-for-one gains achievable in digital network coding for the wireless interference relay channel defined in (\ref{eq:7}). As a simple example, consider  the linear-deterministic channel with modulo-sum interference and a shared out-of-band relay.  Linear-deterministic modulo-sum models represent transmitted and received signals by their corresponding binary expansions, and approximate additions with modulo-sum for simplicity  \cite{adt,adt_2007}.  In many scenarios, linear-deterministic models have been shown to provide useful insights about their corresponding additive-noise channels, e.g. \cite{bresler_tse,adt}.  In Fig.~\ref{fig:det-relay-1}, the relay can assist both users simultaneously by forwarding $a_1\oplus b_1$, which is also the most significant bit (MSB) of the relay observation signal.   Without the relay help, user one and user two can achieve a rate pair of $R_1=1,R_2=2$, by sending $a_1$, and $b_1,b_2$, respectively.  Using the relay message $a_1\oplus b_1$  combined with its own observation $a_1$, the first destination can now decode and cancel interfering bit $b_1$, and thus, the bit $a_2$ can also be delivered to user two.  Similarly, user two can decode interfering bit $a_1$ using relay message $a_1\oplus b_1$ and its own observation $b_1$, allowing to recover an additional bit $b_3$ for user two.  In this case,  the relay can assist both users decode an additional information bit using a single bit.

In the example of Fig.~\ref{fig:det-relay-1}, the key mechanism through which the relay can simultaneously assist both users is providing useful {\em equations} at the receiving ends, evoking again similarities with digital network coding. In the above example, the relay strategy of forwarding $a_1\oplus b_1$ can be interpreted as a quantize-and-forward scheme, with a major difference; unlike conventional quantize-and-forward strategies, here, the purpose of quantize-and-forward at relay is not necessarily to minimize the distortion of relay observation at the destination, but rather to include useful information in the compressed relay message instrumental to decoding of messages at both destinations.

To further see how the relay strategy depends on network configuration, consider now a slightly different network shown in Fig.~\ref{fig:det-relay-2}.  The underlying interference channel in this example is the same as the previous example in Fig.~\ref{fig:det-relay-1}, with the only difference being a weaker interference  from user one at the relay.  Due to weaker interference, the relay now observes $b_1$ free of interference, yet, it is easy to check that a decode-and-forward type of strategy that forwards $b_1$ is not preferred, since user two can always decode $b_1$ on its own without relay's help.  In this case, the relay can enable each source to send one additional bit, $a_3$ and $b_3$, by forwarding a single bit conveying $a_1\oplus b_2$ to the two destinations.  Using $a_1\oplus b_2$ combined with its own observation $a_1$, destination one decodes $b_2$ and subsequently, can recover $a_3$.  Similarly, destination two uses $a_1\oplus b_2$ combined with $b_2$ to decode $a_1$, and then recover $b_3$.  We see that unlike the previous example where it sufficed to quantize the relay observation at the MSB bit level, here, the relay quantization scheme has to accommodate enough resolution to include the second bit level to achieve a two-for-one gain.  Comparing the two relay strategies in above examples reveals that the relay quantization strategy needs to be carefully designed to maximize the gains.

For relay link rate $R_0$  below a certain threshold  that depends on channel parameters, it is shown in this paper that a well-designed form of quantize-and-forward relay strategy achieves the entire capacity region of the Gaussian interference-relay channel defined in (\ref{eq:7}) to within a constant gap of $\Delta=1.95$ bits.  This approximate capacity region also reveals interesting regimes where a two-for-one type of gain is achievable.  Namely, we gain two-to-one for each bit relayed in a weak interference regime which coincides with regimes 1 and 2 ($\alpha<2/3$) identified in  \cite[(25)]{etkin_tse}. Interestingly, the addition of a relay with limited rate does not change the defining boundaries of various interference regimes identified in \cite{etkin_tse}.

A key technique behind the results in this paper is a joint decoding strategy employed at destination ends.  Looking back at examples of Fig.~\ref{fig:det-relay-1} and Fig.~\ref{fig:det-relay-2}, we see that to recover the source message, the decoder needs to solve a linear system of equations.   The benefit of joint decoding has been shown in a number of other contexts, for example, in single-source multiple-relay networks where it is shown that joint decoding is essential to achieve the capacity \cite{peyman_wei_parity_forwarding}.  Yet, an interesting finding  is that in regimes where a two-for-one type of gain is achievable,  a clever successive decoding strategy with an optimal order of decoding source and the relay messages also achieves the two-for-one gain.  However, over the entire capacity region, successive decoding is still suboptimal and results in unbounded gap to capacity, as SNR tends to infinity.

A limiting aspect of the present work is the constraint on $R_0$, the rate of the relay link.  The constraint on $R_0$ arises from the Han-Kobayashi encoding strategy  using obliviously with respect to the relay strategy in place.  For large $R_0$, there are certain regimes in which it is expected that encoding strategy at source nodes should change in presence of the relay node.  To see this, consider the case where the relay rate $R_0$ is unlimited.   In that case, the interference relay channel effectively transforms to a Gaussian interference channel with multiple receive antennas at destination nodes, with correlation between outputs across destinations.   For a MIMO interference channel, it is known that the power splitting between common-private messages used for single-output interference channel is no longer optimal for multiple-antenna destinations \cite{wang_tse}.  Thus, it is expected that, for large values of $R_0$, an encoding strategy oblivious towards relay is not optimal.  Secondly, in certain regimes, random coding strategy at source nodes may not be optimal, and structured codes may be required, as pointed in \cite{ty_2011}.  However, our results show that  for limited rate relays $R_0$, a Han-Kobayashi encoding strategy at source nodes, which is also oblivious towards relay presence, is optimal, to within a constant approximation error. See also  \cite{zhou_incremental_relay} where the approximate capacity of this channel is established in a different weak-relay regime.


\subsection{Related Work}
The interference channel with a relay has been studied under various models in the literature. In \cite{Maric_Dabora}, a two-user interference channel is considered in presence of a relay which observes the signal of only one of the two sources with no interference. For this model, it is shown that although the relay could only observe the signal of one user, it can help the other user also by {\it interference forwarding}, helping the other user subtract interference. In another line of work, a Gaussian linear  interference channel is augmented by a parallel relay channel with incoming and outgoing links orthogonal to the interference channel \cite{sahin_simeone_erkip,sse_2011}. Having dedicated relay links for each user, \cite{sahin_simeone_erkip} and \cite{sse_2011} compare interference forwarding versus signal relaying. The channel model studied in this paper assumes in-band incoming relay links, while the outgoing relay link is {\em shared}. The outgoing broadcast relay link shared between the two destinations is inspired by the broadcast nature of  wireless channels, as motivated before. This channel model studied in this paper was introduced in \cite{peyman_yu_ita10}, where the case of treating interference as noise were considered.   A practical coding strategy for this channel of the case of treating interference as noise may also be found in \cite{rc_2012}.

The Gaussian interference-relay channel with a common relay has been previously treated in \cite{djeumou_belmega_lasaulce,peyman_yu_ita10,noisy_network_coding}. Treating
interference as noise, the classic compress-and-forward (CF) strategy is analyzed in \cite{djeumou_belmega_lasaulce}, where the relay quantizes its observation at a certain resolution so that both destinations reconstruct the relay observation first (see Section~\ref{sec:CF} for discussions on the impact of decoding order). For higher SNR regimes, \cite{peyman_yu_ita10} introduces an improved CF scheme, dubbed generalized hash-and-forward (GHF), following \cite{kim_allerton}.  In \cite{peyman_yu_ita10}, a list decoding strategy is proposed that together with a quantize-and-forward scheme achieves a two-for-one gain for the channel defined in (\ref{eq:7}), when  interference is treated as noise.


\begin{figure}
  \begin{center}
    \includegraphics[width=0.9\columnwidth]{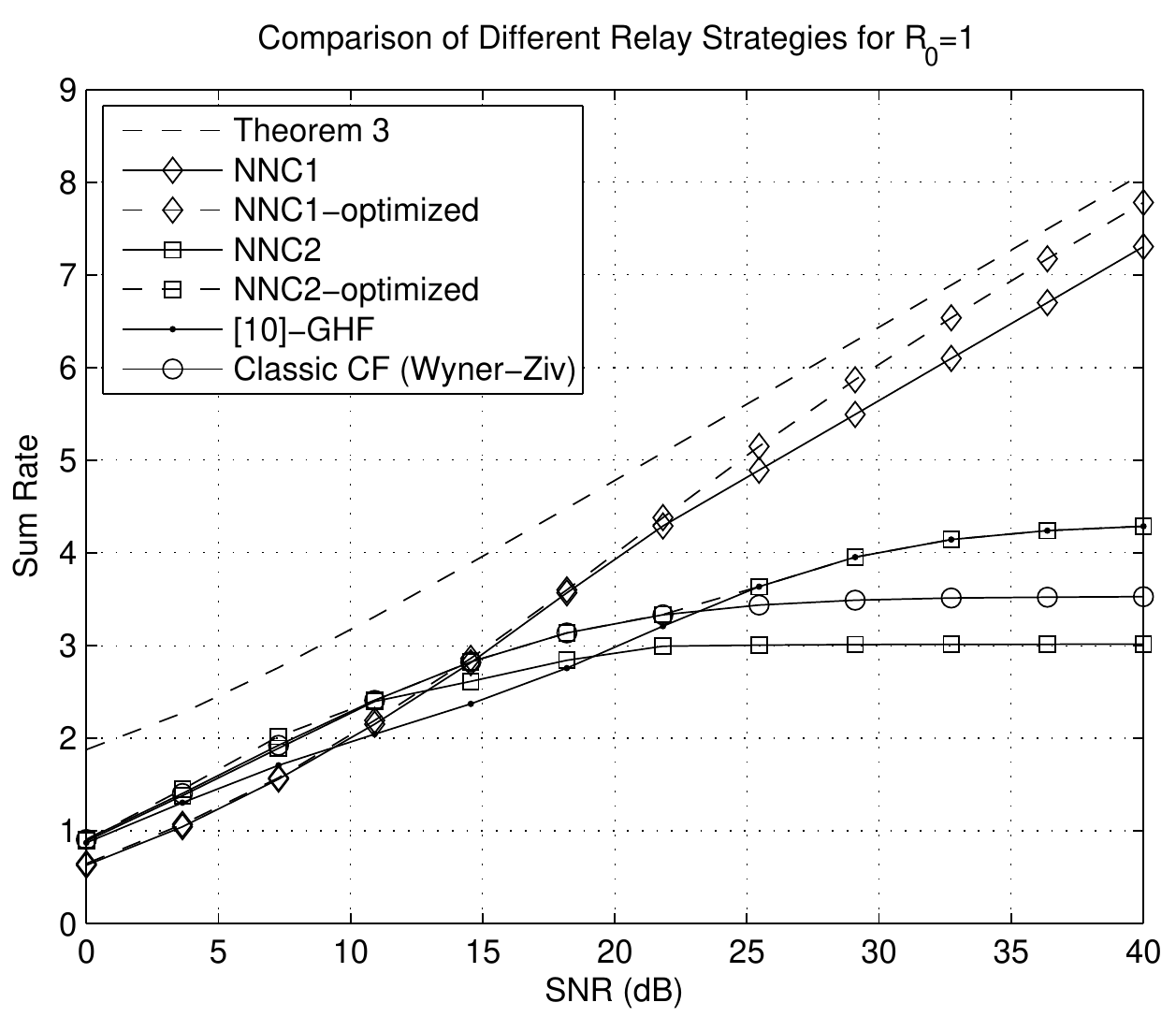}
  \end{center}
  \caption{Comparison of classic CF with Wyner-Ziv encoding and decoding, GHF, and quantize-remap-and-forward strategy in noisy network coding (NNC).  Here, NNC1 and NNC2 correspond to the general noisy network coding strategy of \cite[Section V]{noisy_network_coding} with quantization resolution equal to background noise variance, where interference is completely subtracted (NNC1, \cite[Theorem 2]{noisy_network_coding}) or is treated as noise (NNC2, \cite[Theorem 3]{noisy_network_coding}).  The achievable rates for optimized NNC are obtained by optimizing the quantization resolution, and are the same as results reported in \cite[Fig.~4]{noisy_network_coding}.  The channel parameters are chosen to match those in \cite{noisy_network_coding} with $h_{11}=h_{22}=1,h_{12}=h_{21}=0.5,g_1=0.5,g_2=0.1$, and $N=1$. \label{fig:compare}}
\end{figure}

The channel model studied in this paper is also considered in \cite{noisy_network_coding} in the context of noisy network coding, with a difference of having an analog out-of-band relay link. Noisy network coding employs the quantize-remap-and-forward (QMF) strategy of \cite{adt} with a joint decoding strategy. Two strategies are proposed in \cite{noisy_network_coding}, differing in how interference is handled. When interference is treated as noise, noisy network coding \cite[Theorem~3]{noisy_network_coding}, dubbed NNC2 in Fig.~\ref{fig:compare}, performs similarly as previous CF and GHF schemes of \cite{peyman_yu_ita10}. As shown in Fig.~\ref{fig:compare}, optimized  noisy network coding (with optimization performed over quantization resolution at the relay) achieves the envelope of CF and GHF rates, combined. Theorem~2 of \cite{noisy_network_coding} also considers the case of fully decoding interference, dubbed NNC1 in Fig.~\ref{fig:compare},  which outperforms both GHF scheme of \cite{peyman_yu_ita10} and NNC2 in high SNRs where decoding interference is optimal\footnote{In a Gaussian interference channel, it is optimal to fully decode both signal and interference  when background noise tends to zero while other channel parameters remain constant \cite{etkin_tse}.}.  
Finally, Theorem~\ref{thm:improvement} in this paper improves upon previous strategies as shown in Fig.~\ref{fig:compare}.  A detailed comparison between QMF and other quantize-and-forward strategies is presented in Section \ref{sec:trim-forward-tf}.

\begin{figure}
  \begin{center}
    \includegraphics[width=0.7\columnwidth]{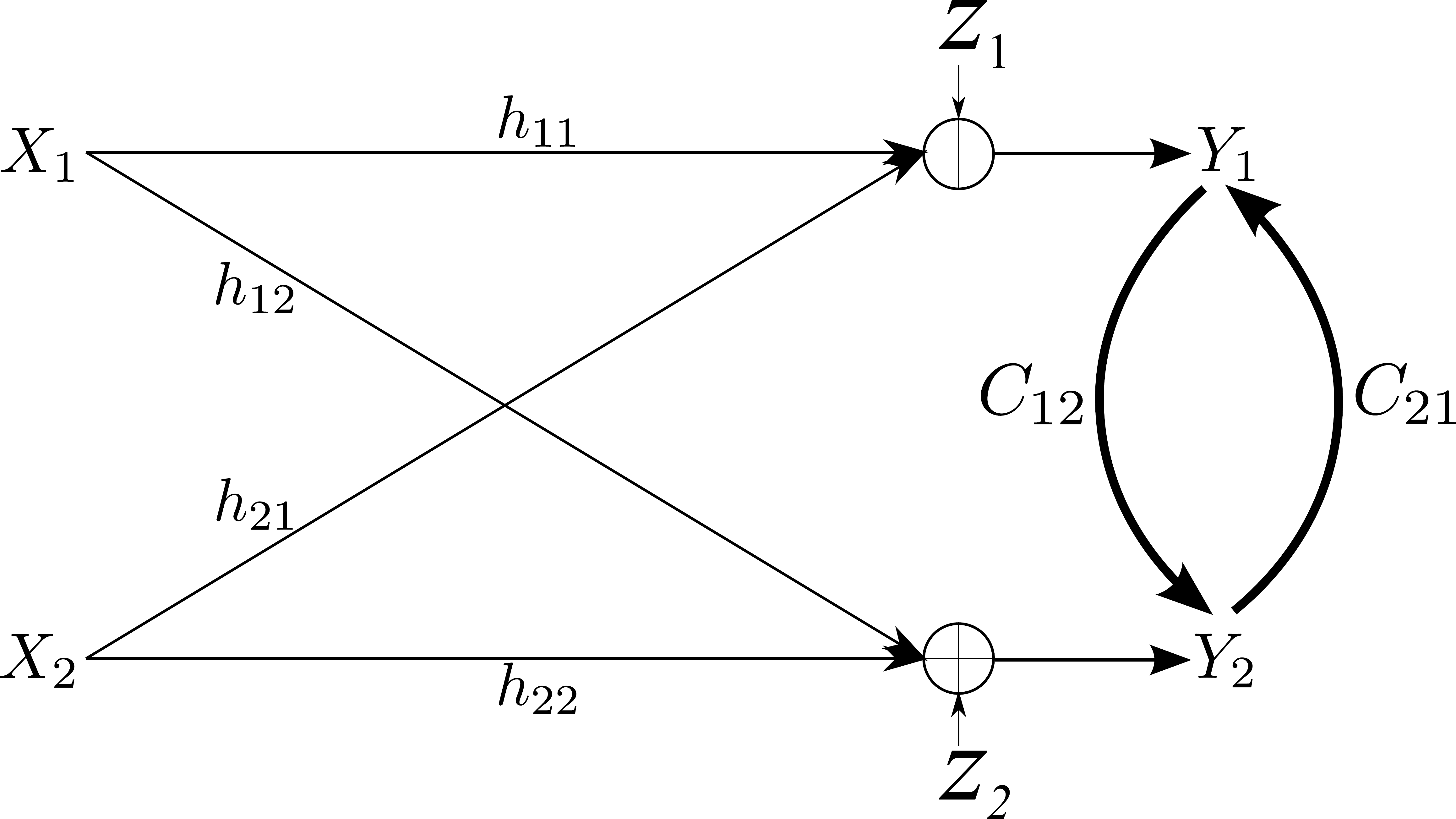}
  \end{center}
  \caption{An interference channel with conferencing receivers. Each receiver
  can also be interpreted as a relay node for the other receiver.}
  \label{fig:conference}
\end{figure}

From another perspective, there is also an interesting connection between the interference relay channel at hand, and an interference channel with conferencing receivers. This channel was first studied in \cite{zhou_yu} for the case of a one-sided interference channel, and a recent comprehensive study is given in
\cite{wang_tse}. Fig.~\ref{fig:conference} shows a linear interference channel
with two conferencing receivers. Each receiver has an out-of-band link of
limited rate to the other receiver. If we only allow for one
simultaneous round of message exchange, we may interpret each destination   as
a relay for the other. For this interference channel with conferencing receivers, QF relay strategies with joint decoding are
considered in \cite{wang_tse}, and the channel capacity region is entirely
characterized to within a constant gap. The issue of choosing the right
quantization level along with appropriate joint decoding, versus employing
successive decoding and conventional Wyner-Ziv type of quantize-and-forward,
also arises for this channel. In \cite{wang_tse}, it is shown that quantizing
the received observation at each user above the power level of the private
messages combined with appropriate number of message exchange rounds is
optimal for this channel. We observe a similar conclusion for the interference relay channel that the relay quantization strategy should be designed to contain only information about common source messages, which are decoded at both destinations.   


\subsection{Organization}
In what follows, the relay quantization strategy along with corresponding decoding scheme  is presented Section~\ref{sec:trim-forward-tf}. Based on this strategy, a general achievable rate is derived for the interference-relay channel in Section~\ref{sec:general}. The achievable rate region is used to characterize the approximate capacity region in Section~\ref{sec:capacity}. Section~\ref{sec:CF} compares quantize-and-forward strategy with joint decoding against a compress-and-forward strategy with successive decoding. Optimal decoding order with successive decoding is also studied in Section~\ref{sec:CF}. 
Finally, Section~\ref{sec:conc} concludes
the paper.


\section{Generalized Hash-and-Forward (GHF) \label{sec:trim-forward-tf}}
Generalized hash-and-forward is  a  quantize-and-forward strategy, where the relay observation is first quantized and then binned
much like conventional compress-and-forward with Wyner-Ziv quantization
\cite[Theorem~6]{cover_elgamal}; the major difference here is that the
quantizer is not constrained to minimize distortion. The decoding strategy in GHF is 
also more general, allowing for more flexible quantization
strategies beyond Wyner-Ziv constraints.

\begin{figure}
  \centering
  \includegraphics[width=0.5\textwidth]{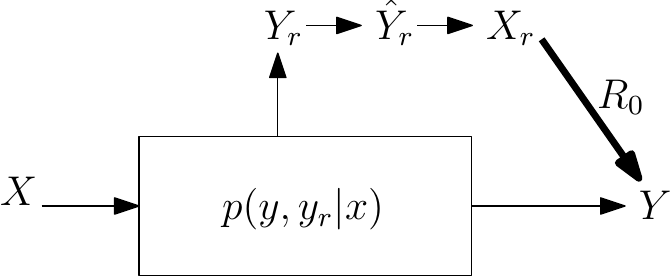}
  \caption{A single-relay channel with an error-free out-of-band relay link of rate $R_0$
  \label{fig:single-relay}}
\end{figure}

Consider a relay channel formed by a source,
a relay, and a destination node, where the relay can communicate to
the destination using a digital link of rate $R_0$, as shown in
Fig.~\ref{fig:single-relay}. Denote the source signal as $X$, and the
relay and destination observations as $Y_r$ and $Y$, respectively.
When the relay cannot decode the source codeword, a sensible relay
strategy is to assist the destination by describing its observation at
rate $R_0$. A central question in the design of
relay strategy is how such quantization should be performed?


In the classic CF scheme \cite[Theorem~6]{cover_elgamal}, the relay
observation is quantized using a Wyner-Ziv source coding technique to minimize distortion at the destination. In this
case, the relay quantizes $Y_r$ using an auxiliary random variable
$\hat{Y_r}$ then sends a bin index at rate $R_0$ to the destination, so that
using side information $Y$, the destination can uniquely recover $\hat{Y_r}$
then proceed to decode $X$ from $Y$ and $\hat{Y_r}$.

Consider now  the more general GHF strategy where we choose an arbitrary
auxiliary random variable $\hat{Y_r}$ to quantize $Y_r$ and provide a bin index
for the quantized codeword to the
destination. Unlike in CF, even with the use of side information
$Y$, the destination can only determine a list $\mathcal{L}$ of possible
quantization relay codewords.  Nevertheless, the destination can still search through
all source codewords by testing the joint typicality of each source
codeword within the list $\mathcal{L}$, then decode a unique $X$.

For the single-relay channel, the above list decoding strategy
gives no higher rate than classic CF. In other words, classic Wyner-Ziv coding is optimal among all GHF strategies, and there is no loss of optimality in restricting the list to be of size 1, i.e., to first decode a unique quantization codeword at the destination. When the relay serves multiple destinations, for example in the relay-interference channel, a single quantization scheme can no longer minimize distortions at multiple destinations at the same time, due to the difference in channel gains and side information. This motivates the use of GHF strategy that allows the flexibility of list decoding at the destinations.

%

 The following theorem yields
the achievable rate of the GHF strategy for an arbitrary relay quantizer
and list decoding at the destination.


\begin{theorem}[Achievable rate of GHF]\label{thm:ghf}\label{thm:2}
  Consider a memoryless single-relay channel defined by $p(y,y_r|x)$,
  where $Y$ and $Y_r$ represent received signals at the destination
  and the relay, with a noiseless (out-of-band) relay link of rate
  $R_0$ bits per channel use. For this channel, the source rate $R$ is achievable if
  \begin{subequations}
  \begin{align}
  R&<\min\bigl\{I(X;Y,\hat{Y}_r),
  I(X;Y)+R_0-I(\hat{Y}_r;Y_r|X,Y)\bigr\}\label{eq:ghfrate-1}\\
&=I(X;Y)+\min\bigl\{R_0,I(\hat{Y_r};Y_r|Y)\bigr\}-I(\hat{Y_r};Y_r|X,Y)\label{eq:midstep}\\
&:=I(X;Y)+\Delta R-\Delta\label{eq:t234}
  \end{align}
  \end{subequations}
  for $(X,Y,Y_r,\hat{Y_r})\sim p(x)p(y,y_r|x)p(\hat{y}_r|y_r)$.
\end{theorem}

\begin{proof} First note that \eqref{eq:midstep} follows from \eqref{eq:ghfrate-1} since we have
\begin{align}
 & \min\bigl\{I(X;Y,\hat{Y}_r),
  I(X;Y)+R_0-I(\hat{Y}_r;Y_r|X,Y)\bigr\}\\
&=I(X;Y)+\min\bigl\{I(X;\hat{Y}_r|Y),R_0-I(\hat{Y}_r;Y_r|X,Y)\bigr\}\NoN\\
&\overset{(a)}=I(X;Y)+\min\bigl\{I(\hat{Y}_r;Y_r|Y)-I(\hat{Y_r};Y_r|X,Y),R_0-I(\hat{Y}_r;Y_r|X,Y)\bigr\}\NoN\\
&=I(X;Y)+\min\bigl\{I(\hat{Y}_r;Y_r|Y),R_0\bigr\}-I(\hat{Y}_r;Y_r|X,Y)\NoN,
\end{align}
where (a) follows since $H(\hat{Y}_r|Y_r,Y)=H(\hat{Y}_r|Y_r,X,Y)$ for the Markov chain $(X,Y)-Y_r-\hat{Y}_r$.

The achievability of the above rate can be proved directly from the CF rate expression in \cite[Theorem~6]{cover_elgamal}, since for a single-relay channel, GHF gives no higher rate than CF, however, CF strategy in \cite[Thoerem~6]{cover_elgamal} cannot be generalized beyond the single-relay channel. Yet, a more general approach based on joint decoding results in the same achievable rate  \cite{dabora_servetto, chong_motani_garg,kim_allerton}.  In  Appendix~\ref{app:proofghf}, a different proof is presented based on list decoding to further illustrate the connections between the classic CF strategy of \cite[Theorem~6]{cover_elgamal} and the more recent strategies based on joint decoding. See also the discussion later in this section.
\end{proof}

\begin{remark}
  The rate improvement due to GHF can be decomposed into two parts, a positive
  improvement $\Delta R$, and a negative penalty $\Delta$.   The negative term
  $\Delta=I(\hat{Y_r};Y_r|X,Y)$ can be interpreted as the
  penalty due to quantization and it is zero if the relay observation $Y_r$
  is a deterministic function of $X$ and $Y$, in which case we say
  $X,Y,Y_r$ form a cross-deterministic relation.  Intuitively, for a relay
  quantizer $\hat{Y}_r$ to be asymptotically cut-set bound achieving, we need
  that   the quantization penalty $\Delta$ tend to zero. We shall see later for
  an interference-relay channel that the quantization penalty of a GHF strategy
  takes a similar form. By choosing a relay quantizer for which the quantization
  penalty is always less than a constant value, we devise a universal relay
  strategy that achieves the capacity of the interference-relay channel to
  within a constant (in the small-$R_0$ regime).
\end{remark}

\subsection{CF, GHF, and Quantize-Map-and-Forward}
The rate expression for GHF in Theorem~\ref{thm:ghf} is identical to the achievable rate of CF, extended-hash-and-forward (EHF), and quantize-map-and-forward  \cite{dabora_servetto, chong_motani_garg,kim_allerton,noisy_network_coding}.
The general encoding strategy in CF, EHF, and GHF is quantization followed by binning, with a more flexible quantization in GHF (and EHF) due to list (or joint)  decoding.  The importance of flexible quantization becomes further clear as we study the interference-relay channel.

\begin{figure}
  \centering
  \includegraphics[width=0.9\textwidth]{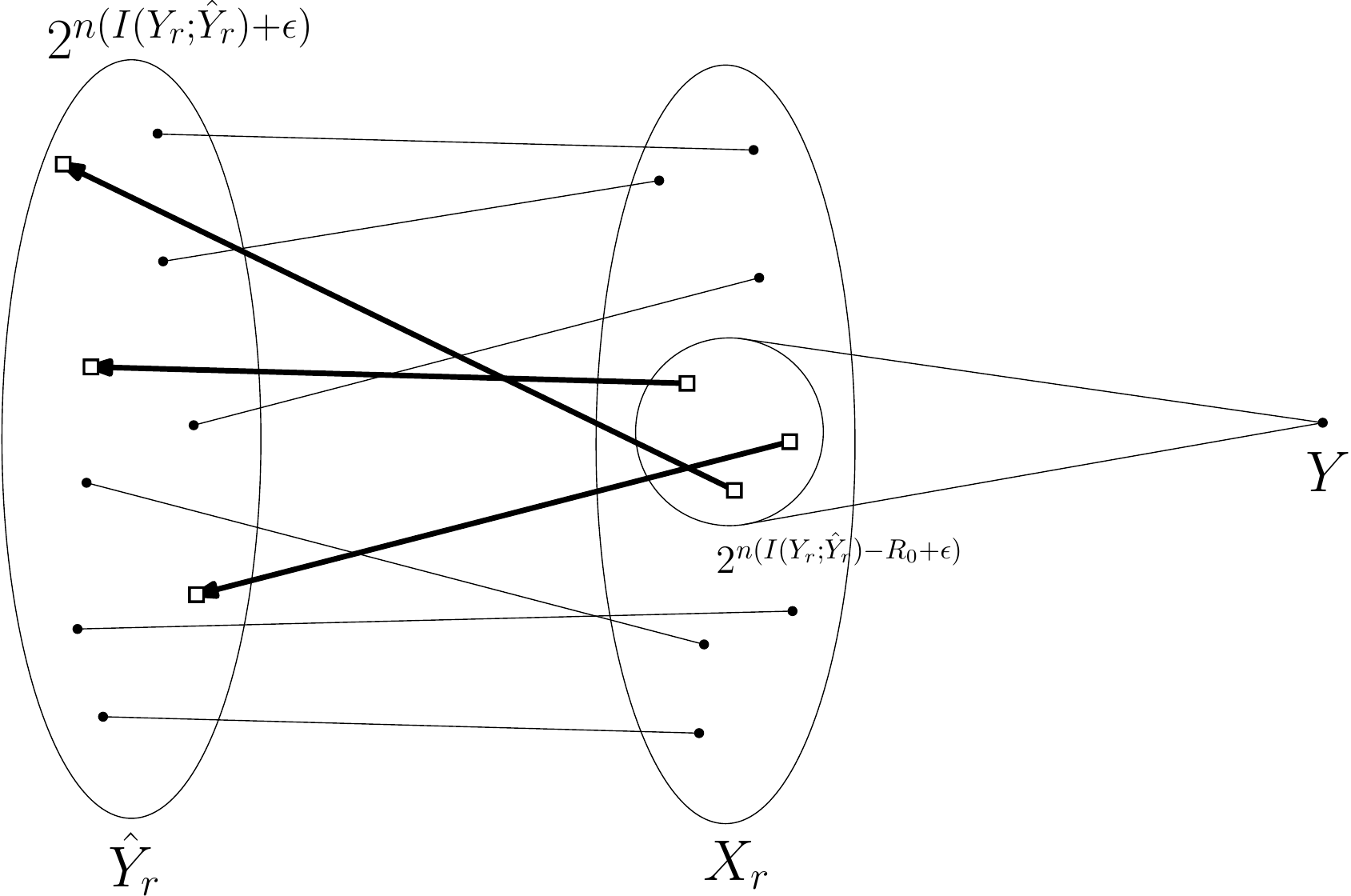}
  \caption{Similarities between Quantize-Map-and-Forward and GHF:
  Since the number of candidate relay codewords that are jointly typical with the received observation $Y^n$ is slashed down by $2^{nR_0}$ asymptotically, the space of candidate quantized relay codewords is also randomly pruned by a $2^{nR_0}$ factor through the random one-to-one mapping between the quantization and channel codes, as if binning automatically occurs at the receiver side.
  \label{fig:QMF}}
\end{figure}

The encoding strategy in QMF is slightly different as compared to CF, since at the first look, there is no explicit use of random binning. Recall that in QMF, the relay employs two codebooks, a quantization code  and a channel code, mapped one-to-one randomly. The relay quantizes its observation using the quantization code, and transmits the corresponding codeword from the channel code. However, a close inspection of QMF reveals the similarities of GHF and QMF: Joint decoding along with random mapping has the same net effect as binning.  This is illustrated in Fig.~\ref{fig:QMF}. Notice that in QMF, the rate of the relay channel code is essentially higher than the relay-destination channel capacity $R_0$, and thus, the destination can narrow its list of candidate relay quantization codewords to a size-$2^{n(I(Y_r;\hat{Y}_r)-R_0+\epsilon)}$ list of codewords. Now, since the number of candidate relay codewords is slashed down by $2^{nR_0}$ asymptotically, the space of candidate quantized relay codewords is also randomly pruned by a $2^{nR_0}$ factor through the random one-to-one mapping between the quantization and channel codes,  as if binning automatically occurs at the receiver side.

By embedding the binning step of GHF into the decoding procedure at the receive side, QMF simplifies the encoding at the relay, which is tremendously helpful in a general network with arbitrary number of relays and possibly loops as in \cite{adt,noisy_network_coding}. However, in terms of actual coding, QMF requires an {\em analog}\footnote{In the sense that the channel allows an input at transmission rate above its capacity.}  channel between the relay and the destination, since otherwise, the relay link cannot be {\em overloaded} above its capacity for the automatic binning to occur at the decoder via joint (or list) decoding. QMF suffers if the relay link is an error-free (digital) bit pipe of limited rate,  since the size of the quantization codebook is then directly constrained by the hard rate limit of the relay link.

\section{An Achievable Rate Region\label{sec:general}}
The GHF strategy can be used along with the common and private message
splitting strategy of Han and Kobayashi (HK) for interference
channel. The resulting achievable rate region is stated in the
following theorem:

\begin{theorem}\label{thm:5}
  For a memoryless interference relay channel defined by
  $p(y_1,y_2,y_r|x_1,x_2)$ with a digital relay link of rate $R_0$
  bits per channel use, a rate pair $(R_1,R_2)$ is achievable if
  $R_1$ and $R_2$ satisfy
  \begin{subequations}\label{eq:relay_HK_simple}
    \begin{align}
      R_1&\leq d_1+\Delta d_1 - \Delta_1\label{eq:r1}\\
      R_2&\leq d_2 +\Delta d_2 -\Delta_2\label{eq:r2}\\
      R_1+R_2&\leq a_1+\Delta a_1- \Delta_1+g_2+\Delta g_2-\Delta_2\label{eq:r3}\\
      R_1+R_2&\leq g_1+\Delta g_1-\Delta_1+a_2+\Delta a_2- \Delta_2\label{eq:r4}\\
      R_1+R_2&\leq e_1+\Delta e_1- \Delta_1+e_2+\Delta e_2-\Delta_2\label{eq:r5}\\
      2R_1+R_2&\leq a_1+\Delta a_1 + g_1+\Delta g_1 -2\Delta_1 + e_2+\Delta e_2  -\Delta_2\label{eq:r6}\\
      R_1+2R_2&\leq  e_1+\Delta e_1 - \Delta_1+a_2+\Delta a_2 + g_2+\Delta g_2 -2\Delta_2 \label{eq:r7}\\
      R_1& \le 0\\
      R_2& \le 0,
    \end{align}
  \end{subequations}
  for some $(Q,W_1,W_2,X_1,X_2,\hat{Y_r},Y_r)\sim
  p(q)p(x_1,w_1|q)p(x_2,w_2|q)p(y_r|x_1,x_2)p(\hat{y}_r|y_r)$, where
  \begin{subequations}\label{eq:a-g}
    \iftwocol
    \begin{align}
      a_1&=I(Y_1;X_1|W_1,W_2,Q)  \\
      d_1&=I(Y_1;X_1|W_2,Q)    \\
      e_1&=I(Y_1;X_1W_2|W_1,Q)  \\
      g_1&=I(Y_1;X_1W_2|Q), \\
      a_2&=I(Y_2;X_2|W_1,W_2,Q)\\
      d_2&=I(Y_2;X_2|W_1,Q)\\
      e_2&=I(Y_2;X_2W_1|W_2,Q)\\
      g_2&=I(Y_2;X_2W_1|Q) \\
      \intertext{and}
      \Delta a_1 &=\min\bigl\{R_0,I(\hat{Y_r};Y_r|Y_1,W_1,W_2,Q)\bigr\}\\
      \Delta d_1&=\min\bigl\{R_0,I(\hat{Y_r};Y_r|Y_1,W_2,Q)\bigr\}\\
      \Delta e_1&=\min\bigl\{R_0,I(\hat{Y_r};Y_r|Y_1,W_1,Q)\bigr\}\\
      \Delta g_1&=\min\bigl\{R_0,I(\hat{Y_r};Y_r|Y_1,Q)\bigr\}\\
      \Delta a_2&=\min\bigl\{R_0,I(\hat{Y_r};Y_r|Y_2,W_1,W_2,Q)\bigr\}\\
      \Delta d_2&=\min\bigl\{R_0,I(\hat{Y_r};Y_r|Y_2,W_1,Q)\bigr\}\\
      \Delta e_2&=\min\bigl\{R_0,I(\hat{Y_r};Y_r|Y_2,W_2,Q)\bigr\}\\
      \Delta g_2&=\min\bigl\{R_0,I(\hat{Y_r};Y_r|Y_2,Q)\bigr\}\\
      \intertext{and}
      \Delta_1&=\min\bigl\{R_0,I(\hat{Y_r};Y_r|Y_1,X_1,W_2,Q\bigr\}\\
      \Delta_2&=\min\bigl\{R_0,I(\hat{Y_r};Y_r|Y_2,X_2,W_1,Q\bigr\}.
    \end{align}
    \else
    \begin{align}
      a_1&=I(Y_1;X_1|W_1,W_2,Q) & \Delta a_1 &=\min\bigl\{R_0,I(\hat{Y_r};Y_r|Y_1,W_1,W_2,Q)\bigr\} \\
      d_1&=I(Y_1;X_1|W_2,Q)   & \Delta d_1&=\min\bigl\{R_0,I(\hat{Y_r};Y_r|Y_1,W_2,Q)\bigr\} \\
      e_1&=I(Y_1;X_1W_2|W_1,Q)  & \Delta e_1&=\min\bigl\{R_0,I(\hat{Y_r};Y_r|Y_1,W_1,Q)\bigr\}\\
      g_1&=I(Y_1;X_1W_2|Q) & \Delta g_1&=\min\bigl\{R_0,I(\hat{Y_r};Y_r|Y_1,Q)\bigr\} \\
      a_2&=I(Y_2;X_2|W_1,W_2,Q) & \Delta a_2&=\min\bigl\{R_0,I(\hat{Y_r};Y_r|Y_2,W_1,W_2,Q)\bigr\}\\
      d_2&=I(Y_2;X_2|W_1,Q)& \Delta d_2&=\min\bigl\{R_0,I(\hat{Y_r};Y_r|Y_2,W_1,Q)\bigr\}\\
      e_2&=I(Y_2;X_2W_1|W_2,Q) & \Delta e_2&=\min\bigl\{R_0,I(\hat{Y_r};Y_r|Y_2,W_2,Q)\bigr\}\\
      g_2&=I(Y_2;X_2W_1|Q) & \Delta
      g_2&=\min\bigl\{R_0,I(\hat{Y_r};Y_r|Y_2,Q)\bigr\}
  \end{align}
  and
\begin{align}
      \Delta_1&=I(\hat{Y_r};Y_r|Y_1,X_1,W_2,Q)\\
      \Delta_2&=I(\hat{Y_r};Y_r|Y_2,X_2,W_1,Q).
    \end{align}
  \fi
  \end{subequations}
\end{theorem}
\begin{proof}
The complete proof is presented in Appendix~\ref{sec:proof}.  The proof is based on combining Han-Kobayashi message splitting technique for the interference channel and the GHF strategy of Theorem~\ref{thm:2}. In Han-Kobayashi message splitting, the source messages are divided into private and common parts encoded using superposition coding. Each destination decodes its own common and private messages, and also the common message of the other user.

Thus, the achievable rate region of  Han-Kobayashi strategy consists of the intersection of the rate regions of two multiple-access channels (MAC).   For this MAC setting, Theorem~\ref{thm:2} can be used to find improvements in the rates of common and private messages. The rate region of the underlying MAC channels are then simplified through a series of eliminations and unions to get the achievable rate region in (\ref{eq:relay_HK_simple}). See Appendix \ref{sec:proof} for details.\end{proof}

It is insightful to compare the Han-Kobayashi rate region in \eqref{eq:relay_HK_simple} with
the rate region of the interference channel without relay. Notice that
the latter takes on the same form of \eqref{eq:relay_HK_simple} without the terms $\Delta a_i,\ldots,\Delta d_i$ and $-\Delta_i$. Therefore, the effect of the relay is to increase each mutual information term $a_i,\ldots,d_i$, by the corresponding quantities $\Delta a_i - \Delta_i,\ldots,\Delta d_i - \Delta_i$, $i=1,2$.
The penalty terms $\Delta_1,\Delta_2$ can be interpreted as the quantization loss. In the next
section, we show that for a Gaussian model in Fig.~\ref{fig:channel-model}, a quantization strategy can be devised to bound the quantization
loss terms $\Delta_1,\Delta_2$ below a constant for all channel coefficients and SNR values. This allows to
prove the achievability of the capacity region of the Gaussian interference channel with an out-of-band relay to within a constant gap, under a constraint on relay link rate.


\section{Approximate Capacity Region in the Weak Interference Regime
\label{sec:capacity}}

Consider a Gaussian interference channel with a digital relay as defined in  \eqref{eq:7}. Following the notation of \cite{etkin_tse}, define
\begin{align}
\SNR_1&:=\frac{P_1\lvert h_{11}\rvert^2}{N}, &  \INR_1&:=\frac{P_2\lvert h_{21}\rvert^2}{N}\NoN \\
\SNR_2&:=\frac{P_2\lvert h_{22}\rvert ^2}{N}, & \INR_2&:=\frac{P_1\lvert h_{12}\rvert^2}{N},\NoN\\
\SNR_{r1}&:=\frac{P_1\lvert g_{1}\rvert ^2}{N}, & \SNR_{r2}&:=\frac{P_2\lvert g_{2}\rvert^2}{N},\NoN
\end{align}
and also let
\iftwocol
\begin{align}
\alpha_1&:=\frac{\log \INR_1}{\log \SNR_1} &
\alpha_2&:=\frac{\log \INR_2}{\log \SNR_2}.
\end{align}
\else
\begin{align}
\alpha_1&:=\frac{\log \INR_1}{\log \SNR_1},&
\alpha_2&:=\frac{\log \INR_2}{\log \SNR_2}\\
\beta_1&:=\frac{\log\SNR_{r1}}{\log\SNR_1},&
\beta_2&:=\frac{\log\SNR_{r2}}{\log\SNR_2}
\end{align}
\fi
We consider the weak interference regime where $0<\alpha_1,\alpha_2<1$. To simplify the derivations, we also assume\footnote{All the derivations can also be performed without this assumption, following exactly the same steps. However, assuming the two direct links are of equal strength is not very limiting and still preserves all interesting regimes of operations.}
\begin{align}
\SNR_1=\SNR_2:=\SNR\NoN.
\end{align}

When $\beta_i,\alpha_i<1, i=1,2$, the following theorem characterizes the capacity
region to within a constant number of bits for a range of values of $R_0$:

\begin{theorem}\label{thm:improvement}
Consider the weak interference regime where $\INR_i< \SNR,i=1,2$. For the case $\SNR_{ri} < \SNR$, a GHF-quantization relay strategy along Han-Kobayashi coding with Etkin-Tse-Wang power splitting strategy achieves the capacity region of the Gaussian interference channel with a common out-of-band relay link of rate $R_0$  to within $1.95$ bits per channel use, for all values of $R_0>0$ satisfying
\begin{subequations}\label{eq:r0cond}
\begin{align}
R_0&\leq \log\SNR+\log \theta\\
R_0&\leq \log\SNR+\log\frac{\INR_2}{\SNR_{r1}}+\log \theta\\
R_0&\leq\log\SNR+\log\frac{\INR_1}{\SNR_{r2}}+\log \theta\\
R_0&\leq \log\frac{\SNR}{\INR_1}+\log\frac{\SNR}{\INR_2}+\log \theta\label{eq:cond4}\\
R_0&\leq\log\frac{\SNR}{\INR_1}+\log\frac{\SNR}{\SNR_{r1}}+\log \theta\\
R_0&\leq\log\frac{\SNR}{\INR_2}+\log\frac{\SNR}{\SNR_{r2}}+\log \theta,
\end{align}
\end{subequations}
where
\begin{align}\NoN
\theta=\min\left\{\norm{\frac{g_1h_{21}-g_2h_{11}}{h_{11}h_{22}}},\norm{\frac{g_2h_{12}-g_1h_{22}}{h_{11}h_{22}}}\right\}.
\end{align}
\end{theorem}
\begin{remark}
The parameter $\theta$ is a measure of dependency between the relay observation $Y_r$ and $Y_1$ and $Y_2$. Notice that $\theta=0$ if either of
\begin{align}
&\left[\begin{array}{cc}g_1 & h_{11}\\g_2 & h_{21}
  \end{array}
\right],&
&\left[\begin{array}{cc}g_1 & h_{12}\\g_2 & h_{22}
  \end{array}
\right]\NoN
\end{align}
is rank deficient, i.e., the relay observation $Y_r$ is statistically equivalent
to $Y_1$ or $Y_2$, in which case the relay can at most reduce noise power  by 3 dBs (through maximal ratio combining).
When $\theta$ is small, the relay can communicate its observation $Y_r$ to the
end users with small errors using side information $Y_1$ and $Y_2$. Thus,
small $\theta$, as well as for large values of $R_0$, the interference relay
channel approximately transforms to a multiple-output interference channel with
$X_1$ and $X_2$ as channel inputs, and $(Y_1,Y_r)$ and $(Y_2,Y_r)$ as channel
outputs. For this channel, Etkin-Tse-Wang power splitting strategy takes a
different form and the power splitting scheme for the single-input single-output
interference channel no longer
achieves the capacity to within a constant, in general; see \cite{wang_tse_mimo_ic}. In other words, when
$R_0$ is large or $\theta$ is small, a different set of strategies are required
to achieve the capacity region. \end{remark}

\begin{proof}
The power splitting strategy of Etkin-Tse-Wang in \cite{etkin_tse}  achieves the capacity region of the underlying interference channel without the relay to within one bit. There, the quantization strategy is designed so that the relay
 signal is quantized at the level of received private messages, or background
 noise, whichever is larger. For this choice of quantization, we see that the capacity is achieved to within a constant gap, when $R_0$ is smaller than a certain threshold.

Using the power splitting strategy of \cite{etkin_tse}, let  $X_1=W_1+V_1$ and $X_2=W_2+V_2$ in Theorem~\ref{thm:5}, where $V_i,W_i$ are independent Gaussian random variables of power $P_{vi}$ and $P_{wi}$, respectively for $i=1,2$, and let
\begin{subequations}\label{eq:etkinsplit}
\begin{align}
P_{v1}&=\frac{N}{h_{12}^2}, &
P_{v2}&=\frac{N}{h_{21}^2},
\intertext{or, equivalently,}
P_{v1}&=\frac{P_1}{\INR_2} & P_{v2}&=\frac{P_2}{\INR_1},
\end{align}
\end{subequations}
i.e., the private message codewords are received at the level of receiver noise. Now, set $\hat{Y_r}=Y_r+\eta$ where $\eta\sim\N(0,q)$ is independent of $Y_r$ and other random variables, and $q$ is given as
\begin{align}\label{eq:qweak}
 q&=\max\left\{N,g_1^2P_{v1},g_2^2P_{v2}\right\}\NoN\\
 &=N\cdot\max\bigl\{1,\SNR_{r1}/\INR_2,\SNR_{r2}/\INR_1\bigr\}
\end{align}
Notice that \eqref{eq:qweak} implies that the relay quantizes its
observation above the power level of private messages and noise. This choice of
$q$ results in a small quantization loss\footnote{Although the quantization
loss is bounded for this quantization level, it may still be not efficient if
the relay rate is above the threshold in \eqref{eq:r0cond}. See an
asymmetric-channel example in Section~\ref{sec:CF}.}, as we have
\begin{align}\label{eq:d1d2}
\Delta_1,\Delta_2<\frac{1}{2}\log \frac{5}{2},
\end{align}
since, for example for $\Delta_1$, we have
\begin{align}
\Delta_1&=I(\hat{Y}_r;Y_r\vert X_1,Y_1,W_2)\NoN\\
&=I(Y_r+\eta;Y_r \vert X_1,Y_1,W_2)\NoN\\
&=I(g_2V_2+Z_r+\eta;g_2V_2+Z_r \vert h_{21}V_2+Z_1)\NoN\\
&=\frac{1}{2}\log\left(1+\frac{N+\norm{g_2}\var\bigl(V_2\vert h_{21}V_2+Z_1\bigl)}{q}\right)\NoN\\
&=\frac{1}{2}\log\left(1+\frac{N}{q}+\frac{\norm{g_2}}{q}\cdot\frac{P_{v2}N}{\norm{h_{21}}P_{v2}+N}\right)\NoN\\
&=\frac{1}{2}\log\left(1+\frac{N}{q}+\frac{\norm{g_2}}{q}\cdot\frac{P_{v2}}{2}\right)\label{eq:delta1}\\
&=\frac{1}{2}\log\left(1+\frac{N}{\max\{N,g_1^2P_{v1},g_2^2P_{v2}\}}+\frac{\norm{g_2}P_{v2}}{2\max\{N,g_1^2P_{v1},g_2^2P_{v2}\}}\right)\NoN\\
&\leq \frac{1}{2}\log\left(1+1+\frac{1}{2}\right)\NoN\\
&= \frac{1}{2}\log\frac{5}{2}\NoN.
\end{align}
This bounds the quantization loss terms $\Delta_1,\Delta_2$ in (\ref{eq:relay_HK_simple}). Next, we show that
\begin{subequations}\label{eq:dgde}
\begin{align}
\Delta e_i&\geq R_0-\frac{1}{2}\log3\\
\Delta g_i&\geq R_0-\frac{1}{2}\log3,
\end{align}
\end{subequations}
for $R_0$ satisfying \eqref{eq:r0cond}. Consider $\Delta e_1$, for which we have
\begin{align}\label{eq:compe1}
\Delta e_1&=\min\{R_0,I(\hat{Y}_r;Y_r\vert Y_1,W_1)\}\NoN\\
\intertext{and,}
&I(\hat{Y}_r;Y_r|Y_1,W_1)=I\Bigl(g_1X_1+g_2X_2+Z_r;\NoN\\
&\qquad g_1X_1+g_2X_2+Z_r+\eta\Big\vert h_{11}X_1+h_{21}X_2+Z_1,W_1\Bigr)\NoN\\
&=\frac{1}{2}\log\left(1+\frac{N}{q}+\frac{\var\bigl(g_1V_1+g_2X_2\lvert h_{11}V_1+h_{21}X_2+Z_1\bigr)}{q}\right)\NoN\\ &=\frac{1}{2}\log\left(1+\frac{N}{q}+\frac{\lvert g_1h_{21}-g_2h_{11}\rvert^2P_{v1}P_2+N(\norm{g_1}P_{v1}+\norm{g_2}P_2)}{q\bigl(\lvert h_{11}\rvert^2P_{v1}+\lvert h_{21}\rvert^2P_2+N\bigr)}\right)\NoN\\
&\geq\frac{1}{2}\log\left(1+\norm{\frac{g_1h_{21}-g_2h_{11}}{h_{11}h_{22}}}\cdot\frac{N\cdot\SNR^2}{q\bigl(\SNR+\INR_1\cdot\INR_2+\INR_2\bigr)}\right)\NoN\\
&\geq\frac{1}{2}\log\left(1+\theta\frac{N\cdot\SNR^2}{q\bigl(\SNR+\SNR^{\alpha_1+\alpha_2}+\SNR^{\alpha_2}\bigr)}\right)\NoN\\
&\geq\frac{1}{2}\log\left(1+\theta\frac{N\cdot\SNR^2}{3q\cdot\max\bigl\{\SNR,\SNR^{\alpha_1+\alpha_2},\SNR^{\alpha_2}\bigr\}}\right)\NoN\\
&=\frac{1}{2}\log\left(1+\theta\frac{\SNR^{2-\max\{1,\alpha_1+\alpha_2\}}}{3q/N\cdot}\right)\NoN\\
&=\frac{1}{2}\log\left(1+\theta\frac{\SNR^{\min\{1,2-\alpha_1-\alpha_2\}}}{3\max\bigl\{N,g_1^2P_{v1},g_2^2p_{v2}\bigr\}/N}\right)\NoN\\
&=\frac{1}{2}\log\left(1+\theta\frac{\SNR^{\min\{1,2-\alpha_1-\alpha_2\}}}{3\max\bigl\{1,\SNR_{r1}/\INR_2,\SNR_{r2}/\INR_1\bigr\}}\right)\NoN\\
&=\frac{1}{2}\log\left(1+\theta\frac{\SNR^{\min\{1,2-\alpha_1-\alpha_2\}}}{3\max\bigl\{1,\SNR^{\beta_1-\alpha_2},\SNR^{\beta_2-\alpha_1}\bigr\}}\right)\NoN\\
&=\frac{1}{2}\log\left(1+\theta\frac{\SNR^{\min\{1,2-\alpha_1-\alpha_2\}}}{3\SNR^{\max\bigl\{0,\beta_1-\alpha_2,\beta_2-\alpha_1\bigr\}}}\right)\NoN\\
&\geq\frac{1}{2}\log\theta+\frac{1}{2}\log\left(\frac{\SNR^{\min\{1,2-\alpha_1-\alpha_2\}}}{\SNR^{\max\{0,\beta_1-\alpha_2,\beta_2-\alpha_1\}}}\right)-\frac{1}{2}\log3\NoN\\
&=\frac{1}{2}\log\theta+\frac{1}{2}\Bigl(
\min\bigl\{1,2-\alpha_1-\alpha_2\}+\min\bigl\{0,\alpha_2-\beta_1,\alpha_1-\beta_2\bigr\}\Bigr)\log\SNR-\frac{1}{2}\log3\NoN\\
&\overset{(a)}\geq R_0-\frac{1}{2}\log3,
\end{align}
where (a) follows from \eqref{eq:r0cond}. Similarly, we can prove that $\Delta e_2>R_0-0.5\log3$.

Now, it is proved in Appendix~\ref{sec:proof}, \eqref{eq:dddg},  that $\Delta g_i\geq
\Delta e_i$. Hence, we also have
\begin{align}\label{eq:dg}
\Delta g_i\geq R_0-\frac{1}{2}\log3,
\end{align}
for $i=1,2$.

By Theorem~\ref{thm:5} and \eqref{eq:dgde} and \eqref{eq:d1d2}, we find that
the following rate region is achievable for $R_0$ satisfying~\eqref{eq:r0cond}:
\begin{align}\label{eq:rateregion}
      R_1&\leq d_1\NoN\\ \NoN
      R_2&\leq d_2\\\NoN
      R_1+R_2&\leq a_1+g_2+R_0-\frac{1}{2}\log3-\frac{1}{2}\log\frac{5}{2}\\\NoN
      R_1+R_2&\leq g_1+a_2+R_0-\frac{1}{2}\log3-\frac{1}{2}\log\frac{5}{2}\\\NoN
      R_1+R_2&\leq e_1+e_2+2R_0-\log3-\log\frac{5}{2}\\\NoN
      2R_1+R_2&\leq a_1 + g_1+ e_2+2R_0-\log3-\log\frac{5}{2}\\\NoN
      R_1+2R_2&\leq  e_1 +a_2+ g_2+2R_0-\log3-\log\frac{5}{2} \\\NoN
      R_1&>0\\
      R_2&>0,
\end{align}
where $a_i,d_i,e_i,g_i$ are computed for Etkin-Tse-Wang power splitting strategy with $W_i,X_i$ given in (\ref{eq:etkinsplit}).

To find the gap between the above region and the capacity, Theorem ~\ref{thm:boundsweak} and Corollary 2 in Appendix \ref{sec:bounds} give the  following upper bound for the capacity region of the Gaussian interference channel with an out-of-band relay link of rate $R_0$, when $\SNR_{ri}\leq \SNR$, $i=1,2$:
\begin{align}\NoN
      R_1&\leq d_1+1\\\NoN
      R_2&\leq d_2 +1 \\\NoN
      R_1+R_2&\leq a_1+g_2+R_0+\frac{3}{2}\\\NoN
      R_1+R_2&\leq g_1+a_2+R_0+\frac{3}{2}\\\NoN
      R_1+R_2&\leq e_1+e_2+2R_0+1\label{eq:bounduseful}\\
      2R_1+R_2&\leq a_1 + g_1+ e_2+2R_0+2\\\NoN
      R_1+2R_2&\leq  e_1 +a_2+ g_2+2R_0+2 \\\NoN
      R_1&>0\\\NoN
      R_2&>0,\NoN
\end{align}
where again $d_i,a_i,g_i,e_i$ are computed for Etkin-Tse-Wang power splitting given in \eqref{eq:etkinsplit}. Comparing the outer-bound and the achievable region, we find that the achievable rate region using  GHF relay strategy combined with Etkin-Tse-Wang power splitting is within $0.5\log 15$ bits of the capacity region. This proves the theorem.
\end{proof}


\subsection{Asymptotic Sum Rate Improvement}
From \eqref{eq:rateregion} we observe that a relay link of rate $R_0$  improves the sum rate  by approximately either $2R_0$ or $R_0$ bits per channel use, for constrained $R_0$.  Whether the gain in sum rate is $R_0$ or $2R_0$ depends on the active constraints in \eqref{eq:rateregion}. This section identifies these regimes asymptotically as SNR tends to infinity.

To analyze the asymptotic sum rate, let $R_0=0.5\rho\cdot\log\SNR$ and let $\SNR$ tend to infinity for fixed $\beta_i,\alpha_i,\rho$.  First, we find asymptotic first-order expansions for $a_i,g_i,e_i$ as $\SNR,\INR_i\ra \infty$ for fixed $\alpha_i$.  As $\SNR,\INR_i\ra \infty$, we have:
\begin{subequations}\label{eq:d-e-limits}
\begin{align}
d_1&=I(X_1;Y_1|W_2)\NoN\\
&\ra \frac{1}{2}\log\SNR+O(1),
\intertext{and}
a_1&=I(Y_1;X_1|W_1,W_2)\NoN\\
&=\frac{1}{2}\log\left(1+\frac{\norm{h_{11}}P_{v1}}{\norm{h_{21}}P_{v2}+N}\right)\NoN\\
&=\frac{1}{2}\log\left(1+\frac{\SNR}{2\INR_2}\right)\NoN\\
&\ra  \frac{1}{2}\log\left(\frac{\SNR}{\INR_2}\right)+O(1)\NoN\\
&=\frac{1}{2}(1-\alpha_2)\log\SNR+O(1),\label{eq:a1limit}\\
\intertext{and}
g_1&=I(Y_1;X_1,W_2)\NoN\\
&=I(X_1;Y_1)+I(W_2;Y_1|X_1)\NoN\\
&=\frac{1}{2}\log\left(1+\frac{\SNR}{1+\INR_1}\right)+\frac{1}{2}\log\left(\frac{1+\INR_1}{2}\right)\NoN\\
&\ra \frac{1}{2}\log\left(\SNR\right)+O(1)\label{eq:g1limit}.\\
\intertext{Similarly, we have:}
e_1&=I(Y_1;X_1,W_2|W_1)\NoN\\
&=I(Y_1;X_1|W_1)+I(Y_1;W_2|X_1,W_1)\NoN\\
&=\frac{1}{2}\log\left(1+\frac{\SNR}{\INR_2\cdot(1+\INR_1)}\right)+ \frac{1}{2}\log\left(\frac{1+\INR_1}{2}\right)\NoN \\
&\ra \frac{1}{2}\log\left(\INR_1+\frac{\SNR}{\INR_2}\right)+O(1)\NoN\\
&= \frac{1}{2}\max\left\{\log \INR_1,\log\frac{\SNR}{\INR_2}\right\}+O(1)\NoN\\
&=\frac{1}{2}\max\{\alpha_1,1-\alpha_2\}\log\SNR+O(1)\label{eq:e1limit}
\end{align}
\end{subequations}
Switching  indices, we also obtain asymptotic first-order expansions for $d_2,a_2,g_2,e_2$.

Now, using \eqref{eq:rateregion}  and \eqref{eq:d-e-limits} and neglecting first order terms,  we get the following asymptotic first-order expansion for the rate region for $R_0$ satisfying \eqref{eq:r0cond}:
\begin{subequations}
\begin{align}
R_1&\leq \frac{1}{2}\log \SNR\NoN\\
R_2&\leq \frac{1}{2}\log \SNR\\
R_1+R_2&\leq\frac{1}{2} (2-\alpha_1)\log\SNR+R_0\NoN\\
R_1+R_2&\leq \frac{1}{2}(2-\alpha_2)\log\SNR+R_0\NoN\\
R_1+R_2&\leq \frac{1}{2}\max\bigl\{\alpha_1+\alpha_2,2-\alpha_1-\alpha_2\bigr\}\log\SNR+2R_0\\
2R_1+R_2&\leq  \frac{1}{2}\bigl(2-\alpha_2+\max\{\alpha_2,1-\alpha_1\}\bigr)\log\SNR +2R_0\label{eq:aaa}\\
R_1+2R_2& \leq \frac{1}{2}\bigl(2-\alpha_1+\max\{\alpha_1,1-\alpha_2\}\bigr)\log\SNR+2R_0 \label{eq:bbb},
\end{align}
\end{subequations}
which gives the following constraints on the asymptotically-achievable sum rate:
\begin{subequations}\label{eq:sumconstraints}
\iftwocol
\begin{align}
R_1+R_2&\leq\frac{1}{2} (2-\alpha_1)\log\SNR+R_0\label{eq:con1}\\
R_1+R_2&\leq \frac{1}{2}(2-\alpha_2)\log\SNR+R_0
\label{eq:con2}\\
R_1+R_2&\leq \frac{1}{2}\max\{\alpha_1+\alpha_2,2-\alpha_1-\alpha_2\}\log\SNR\NoN\\
&\qquad \qquad \qquad \qquad \qquad \qquad \quad+2R_0\label{eq:con3}
\end{align}
\else
\begin{align}
R_1+R_2&\leq\frac{1}{2} (2-\alpha_1)\log\SNR+R_0\label{eq:con1}\\
R_1+R_2&\leq \frac{1}{2}(2-\alpha_2)\log\SNR+R_0
\label{eq:con2}\\
R_1+R_2&\leq \frac{1}{2}\max\{\alpha_1+\alpha_2,2-\alpha_1-\alpha_2\}\log\SNR+2R_0\label{eq:con3}
\end{align}
\fi
\end{subequations}
From \eqref{eq:sumconstraints}, we distinguish two different regions for the sum-rate improvement. When $\alpha_1+2\alpha_2<2$ and $2\alpha_1+\alpha_2<2$, \eqref{eq:con3} is the active constraint and every bit relayed improves the sum rate by two bits, asymptotically; otherwise, we get one bit improvement in sum rate, for every bit relayed.

\begin{figure}
\centering
\includegraphics[scale=1]{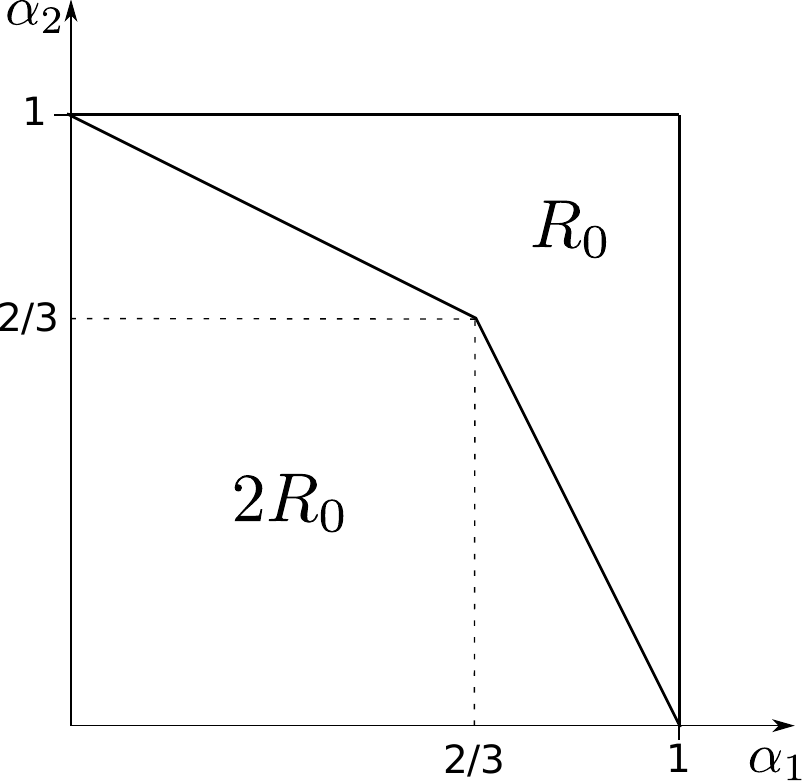}
\caption{The asymptotic improvement in sum-rate with a common digital relay link of rate $R_0$ for different values of $\alpha_1,\alpha_2$ in the weak interference regime. \label{fig:alpha}}
\end{figure}

However, Theorem~\ref{thm:improvement}  only holds for $R_0$ satisfying \eqref{eq:r0cond}. We can further express \eqref{eq:r0cond} in terms of constraints on $\rho,\alpha_i,\beta_i$ in the asymptotic case. To this end, first note that:
\begin{align}
\theta_1&=\norm{\frac{g_1h_{21}-g_2h_{11}}{h_{11}h_{22}}}\NoN\\
&=\norm{\frac{g_1h_{21}\sqrt{P_1P_2}-g_2h_{11}\sqrt{P_1P_2}}{h_{11}h_{22}\sqrt{P_1P_2}}}\ra \SNR^{\max\{\beta_1+\alpha_1-2,\beta_2-1\}} \\
\theta_2&=\norm{\frac{g_2h_{12}-g_1h_{22}}{h_{11}h_{22}}} \ra \SNR^{\max\{\beta_2+\alpha_2-2,\beta_1-1\}}
\intertext{and}
\theta&=\min\{\theta_1,\theta_2\}=\SNR^\tau\NoN,\\
\intertext{where}
\tau&=\min\Bigl\{\max\{\beta_1+\alpha_1-2,\beta_2-1\},\max\{\beta_2+\alpha_2-2,\beta_1-1\}\Bigr\},
\end{align}
asymptotically as $\SNR\ra\infty$.

Thus, we have the following constraints on $\rho$ for \eqref{eq:r0cond} to hold asymptotically:
\begin{align}\NoN
\rho&\leq 1+\tau\\\NoN
\rho&\leq 1+\alpha_2-\beta_1+\tau\\\NoN
\rho&\leq 1+\alpha_1-\beta_2+\tau\\\NoN
\rho&\leq 2-\alpha_1-\alpha_2+\tau\\\NoN
\rho&\leq 2-\alpha_1-\beta_1+\tau\\
\rho&\leq 2-\alpha_2-\beta_2+\tau.
\end{align}
To simplify, consider the case where $\SNR\ra\infty$ while $\beta_1,\beta_2\ra 1^{-}$ for fixed $\alpha_1,\alpha_2<1$.  This asymptotic scenario corresponds to the Gaussian interference-relay channel in (\ref{eq:7}) with fixed $\abs{h_{11}},\abs{h_{22}},\abs{g_1},\abs{g_2}\gg \abs{h_{12}},\abs{h_{21}}\gg N$ as $N\ra 0$. For $\beta_1,\beta_2\ra 1^{-}$ (tend to one from below), the above constraints on $\rho$ reduce to
\begin{align}\label{eq:rhosimple}
\rho\leq \min\{\alpha_1,\alpha_2,1-\alpha_1,1-\alpha_2\}.
\end{align}

Now, from \eqref{eq:sumconstraints},  the sum rate is asymptotically improved by
$2R_0$ bits when $\alpha_1+2\alpha_2<2$ and $\alpha_2+2\alpha_1<2$, and by $R_0$
bits otherwise. We see that for large $R_0$ satisfying \eqref{eq:rhosimple} (implying the constraint in \eqref{eq:r0cond} in asymptotic sense), we get $2$ or $1$ bits of improvement per bit  relayed depending on whether $0<\alpha_1,\alpha_2<1$, and $\alpha_1+2\alpha_2<2$ and $\alpha_2+2\alpha_1<2$ or not, asymptotically. The sum-rate improvement in different regimes is shown in Fig.~\ref{fig:alpha}.

\section{Comparison with Conventional CF\label{sec:CF}}
In this section, the achievable region of Theorem~\ref{thm:improvement} with the region achievable by a conventional CF strategy based on Wyner-Ziv source coding with successive decoding is studied.
Although CF requires a successive decoding strategy, different decoding orders are possible. Since there are
two common messages and a private message to decode, with the addition of the
relay codeword, the end receiver would have four messages to decode. The
messages decoded first assist the decoding of the remaining messages as side
information. Thus, the question of choosing the optimal decoding order for
CF is inevitable: Should the destination first reconstruct the quantized relay codeword, and then use it to decode the two common messages and the private message, or should the decoder first decode for example its own common message, and then reconstruct the quantized relay codeword to finally decode the remaining part of source message? The answer to this question also clarifies how the relay is effectively helping in the GHF strategy with joint decoding.

\subsection{Successive Decoding with Decoding  Relay Quantized Observation First \label{sec:decoderelayfirst}}
A natural decoding strategy is to first reconstruct the relay observation, and use the relay observation to help with decoding of other messages. To reconstruct the relay observation, the destinations use their own observation as side information and the relay performs Wyner-Ziv source coding. Wyner-Ziv quantization with $\hat{Y}_r$ requires that
\begin{align}
 R_0\geq \max\bigl\{ I(\hat{Y_r};Y_r|Y_1), I(\hat{Y_r};Y_r|Y_2)\bigr\}\NoN
\end{align}
for an auxiliary random variable $\hat{Y}_r$. For $\hat{Y}_r=Y_r+\eta$ with $\eta\sim\N(0,q)$, the above constraints give the following value for $q$:
\begin{align}\label{eq:qcf}
q&=\frac{1}{2^{2R_0}-1}\max\bigl\{\var(Y_r|Y_1),\var(Y_r|Y_2)\bigr\}.
\end{align}
The achievable rate using  this quantization strategy is computed in
Appendix~\ref{sec:cf1}.

\subsection{Successive Decoding with Decoding Common Message first}
In this case, the decoded common  message serves as additional side information to reconstruct the relay observation. This would be a reasonable strategy in a moderately-weak ($1/2<\alpha<2/3$) interference regime, where the channel strength over the direct channel is larger than the one over the interference link. Thus, the user can safely decode its own common message with no help from the relay, since it is the cross channel that constrains the rate of the common messages in this regime.

Once $W_1^n$ at user 1 and $W_2^n$ at user 2 are decoded, the relay can use
Wyner-Ziv source coding to communicate its quantized codeword to both
destinations. Decoding  is successful, if
\begin{align}\label{eq:cf2wynerziv}
R_0&\geq \max\left\{I(Y_r;\hat{Y}_r\big\vert
  W_1Y_1),I(Y_r;\hat{Y}_r\big\vert
  W_2Y_2)\right\}\NoN\\
  &=\frac{1}{2}\max\left\{\log\left(1+\frac{\var(Y_r\vert Y_1W_1)}{q}\right),\log\left(1+\frac{\var(Y_r\vert Y_2W_2)}{q}\right)\right\}
\end{align}
for $\hat{Y}_r=Y_r+\eta$ with $\eta\sim\N(0,q)$.

To satisfy \eqref{eq:cf2wynerziv},
the relay quantizes its observation $Y_r$ using $\hat{Y}_r=Y_r+\eta$ where  $q$
is given as
\begin{align}\label{eq:qgoodcf}
  q=\frac{1}{2^{2R_0}-1}\max\bigl\{\var(Y_r\vert Y_1W_1),\var(Y_r\vert
  Y_2W_2)\bigr\}.  \end{align}
The resulting rate is computed in Appendix~\ref{sec:cf2}.

\subsection{Comparison with GHF}
Is there an advantage in  GHF as compared to CF, and if any, under what conditions? To answer this question,  two asymptotic scenarios are studied in this section.

Consider an asymptotic scenario where $\SNR\ra\infty$ while $\beta_1,\beta_2\ra 1^{-}$ for fixed $\alpha_1,\alpha_2<1$.  This asymptotic scenario corresponds to the Gaussian interference-relay channel in (\ref{eq:7}) with fixed $\abs{h_{11}},\abs{h_{22}},\abs{g_1},\abs{g_2}\gg \abs{h_{12}},\abs{h_{21}}\gg N$ as $N\ra 0$.

\subsubsection{Symmetric Case}  In the symmetric case, we have
$\alpha_1=\alpha_2=\alpha$. From (\ref{eq:rhosimple}) and
\eqref{eq:sumconstraints}, we can prove that GHF with quantization strategy of
(\ref{eq:qweak}) gives the following asymptotic achievable sum rate (see
\eqref{eq:app:ghf} of Appendix~\ref{sec:app:cf}):
\begin{subequations}\label{eq:sum-sym-ghf-1}
\begin{align}
  R_1+R_2&\leq\frac{1}{2}
  (2-\alpha)\log\SNR+\frac{1}{2}\min(\rho,\alpha)\log\SNR\\ R_1+R_2&\leq
  \max(\alpha,1-\alpha)\log\SNR+\min(\rho,\alpha,1-\alpha)\log\SNR.
\end{align}
\end{subequations}
Note that the above achievable sum rate in general holds for $R_0$ values beyond
the constraints in \eqref{eq:r0cond}. If the constraints in \eqref{eq:r0cond} are violated,
GHF still gives an achievable rate region, although the same
constant-gap-to-capacity result  may not apply.

Now for this asymptotic scenario, it is proved in  \eqref{eq:app:cf1} of Appendix~\ref{sec:app:cf} that when the relay observation is first reconstructed, the  symmetric achievable sum rate using CF is given by:
\begin{align}\label{eq:cf1sumrate}
R_1+R_2&\leq\frac{1}{2} (2-\alpha)\log\SNR+\frac{1}{2}\min(\rho,\alpha)\log\SNR\NoN\\
R_1+R_2&\leq \max(\alpha,1-\alpha)\log\SNR+\Bigl(\rho+1-\max(1,2\alpha)\Bigr)^+\log\SNR-\Bigl(\rho-\alpha\Bigr)^+\log\SNR
\end{align}
By considering the other decoding order where each user first decodes its own common message, we get the following achievable rate with CF (see Appendix \ref{sec:app:cf}):
\begin{align}\label{eq:cf2sumrate}
R_1+R_2&\leq 2(1-\alpha)\log\SNR\NoN\\
R_1+R_2&\leq \max(\alpha,1-\alpha)\log\SNR+\rho\log\SNR-\Bigl(\rho-\min(\alpha,1-\alpha)\Bigr)^+\log\SNR
\end{align}
The achievable rate using CF is then given as the maximum of the two decoding orders.

\begin{figure}
\centering
\subfigure[]{\includegraphics[height=10cm]{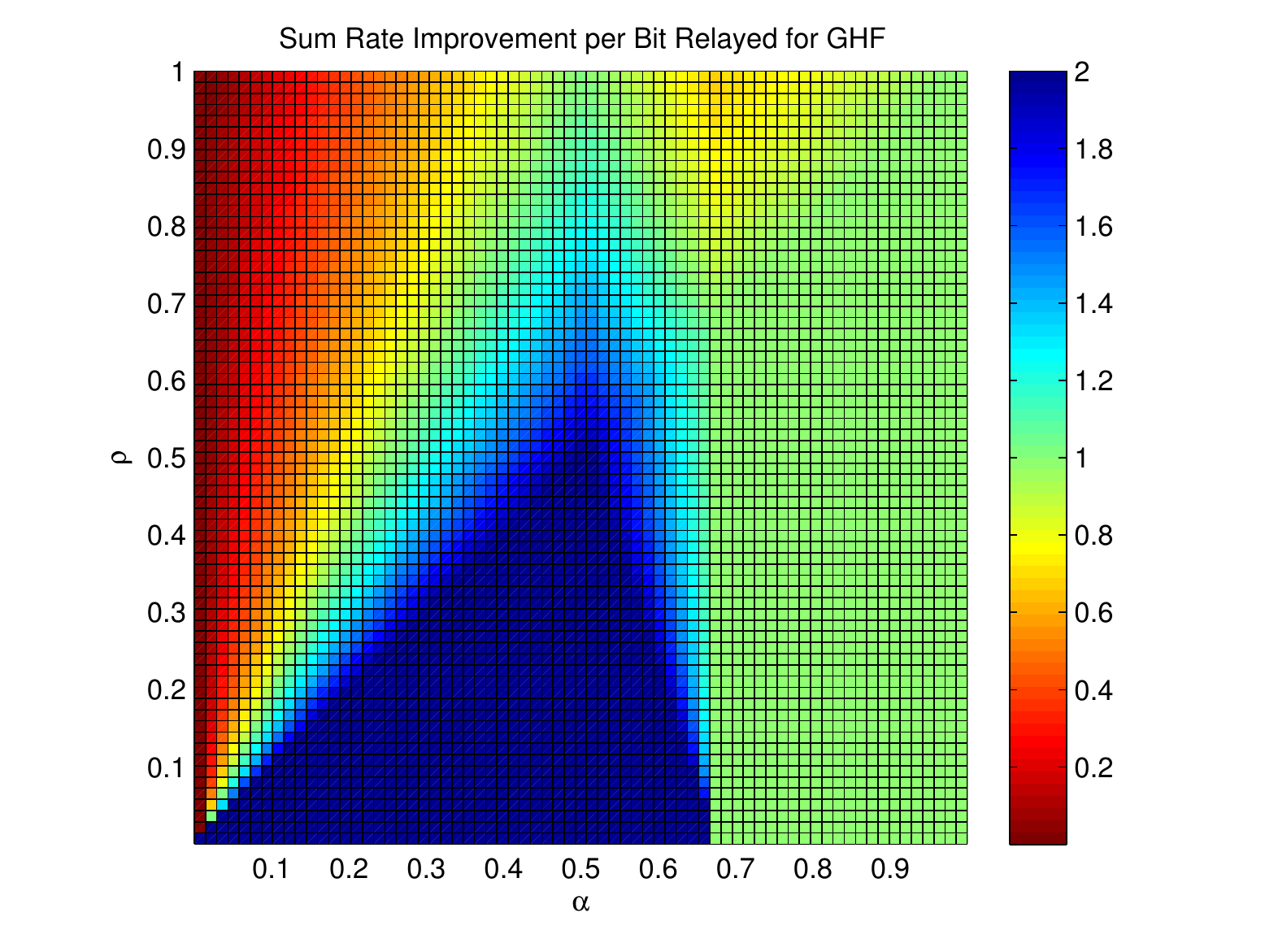}}
\subfigure[]{\includegraphics[height=10cm]{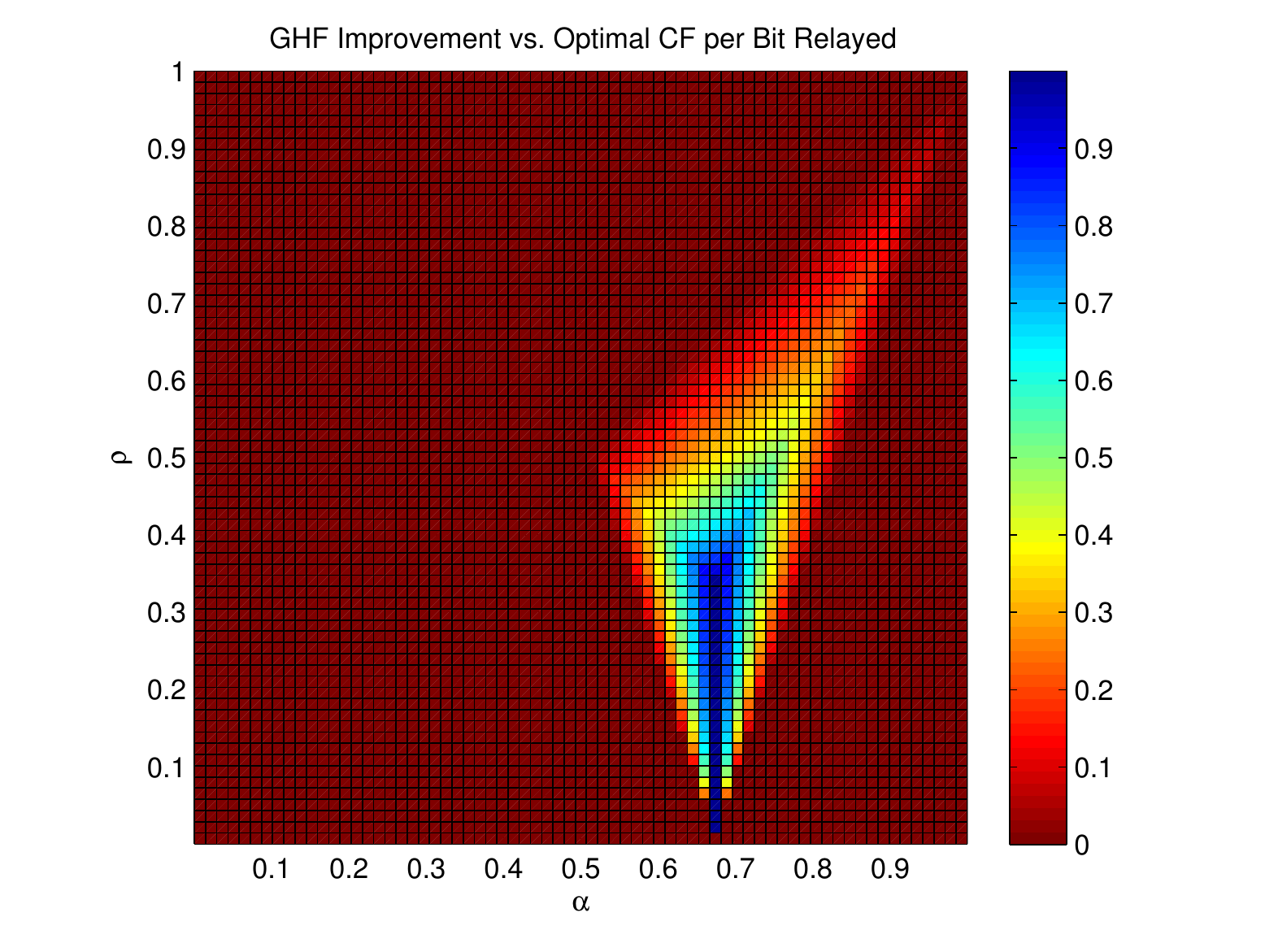}}
\caption{Asymptotic comparison between GHF and CF in symmetric case. (a) Asymptotic sum rate improvement per 1 bit relayed using GHF.  (b) GHF improvement upon CF with two possible decoding orders, per bit relayed. GHF significantly outperforms CF around $\alpha=2/3$. \label{fig:hkghf}}
\end{figure}

Fig.~\ref{fig:hkghf} shows a comparison between CF and GHF sum rates in the
asymptotic regime. The figure shows the asymptotic rate improvement for every
bit relayed for different values of $\rho$ and $\alpha$ in a symmetric
interference-relay channel. As shown in Fig.~\ref{fig:hkghf}-(a) for GHF, when
$\alpha\leq\rho,\rho+3\alpha\leq 2$, we gain asymptotically 2 bits improvements in sum
rate for every bit relayed. For $\rho<\alpha$ and when $\rho+3\alpha> 2$, the
gain in sum rate per bit relayed decreases. In particular, for $\alpha>2/3$, we
asymptotically have only one bit of improvement per relayed bit. For
$\rho>\alpha$, the gain in sum rate with GHF is independent of $\rho$, or
equivalently $R_0$ as \eqref{eq:sum-sym-ghf-1} shows.

\begin{figure}
\centering
\subfigure[]{\includegraphics[height=10cm]{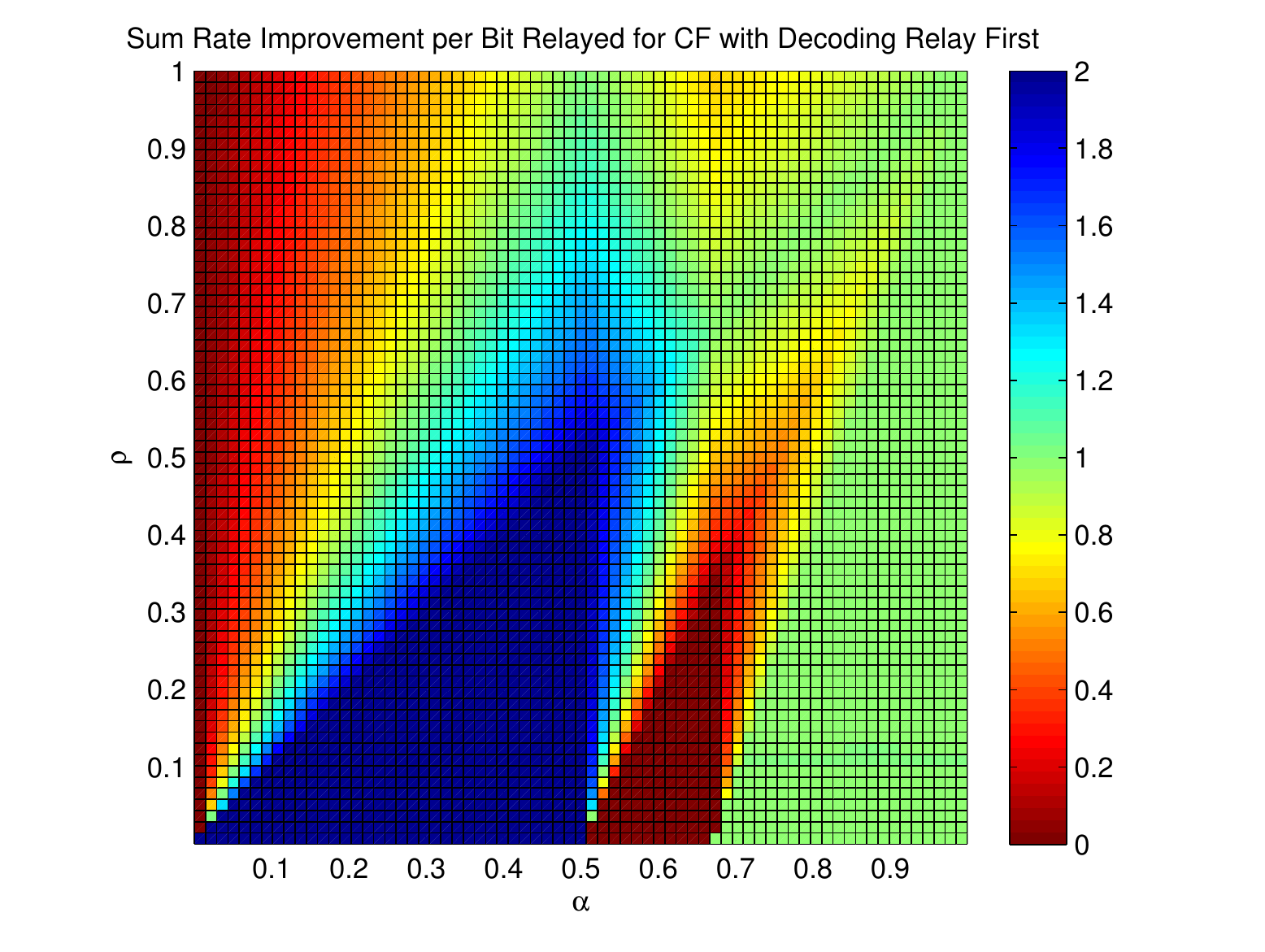}}
\subfigure[]{\includegraphics[height=10cm]{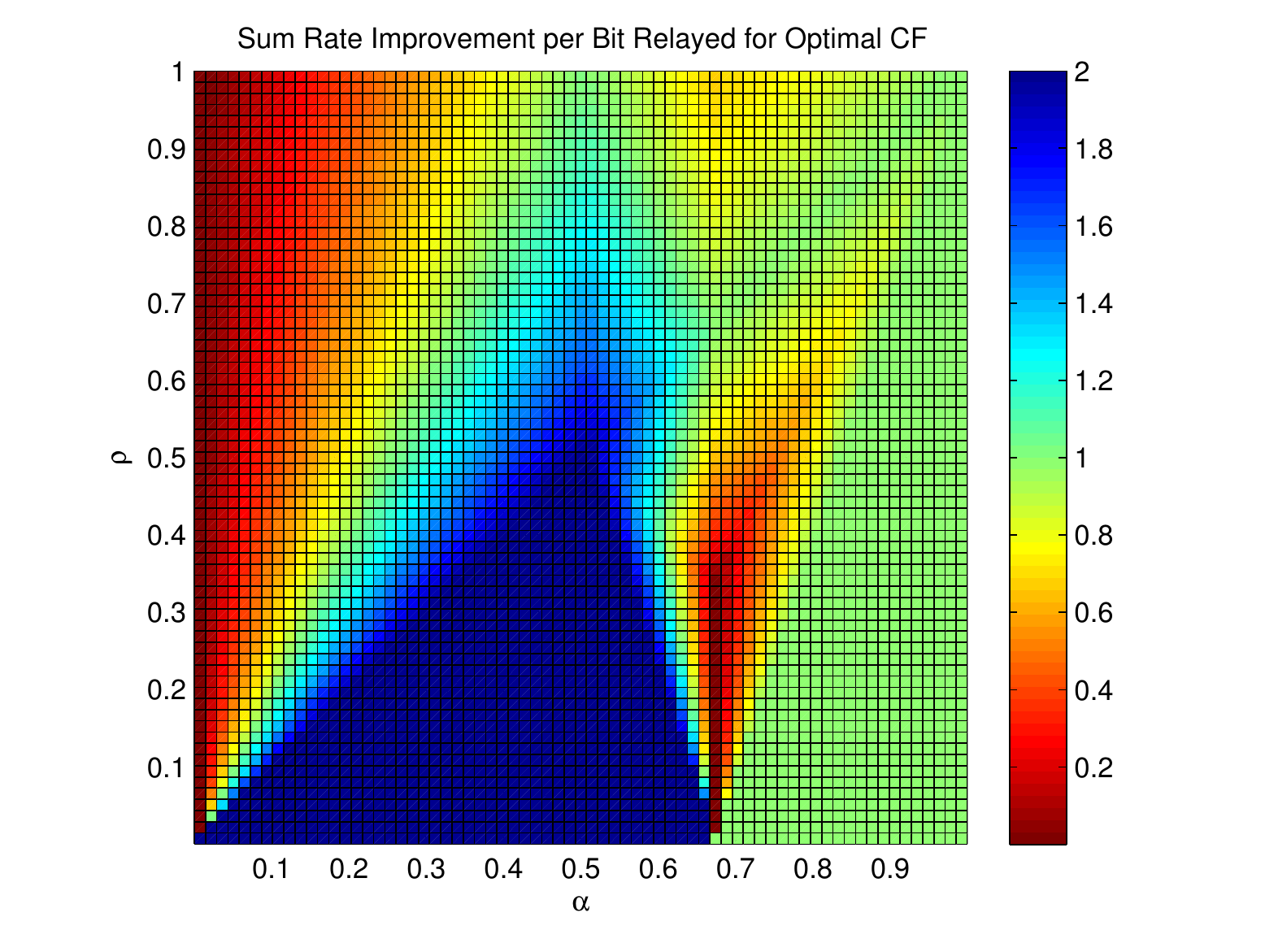}}
\caption{Asymptotic comparison of CF with different decoding orders symmetric interference channel. In (a), the relay observation is always reconstructed first. In (b), optimal decoding order is used. Reconstructing the relay observation first provides no asymptotic gain for $1/2<\alpha<2/3,\rho<2\alpha-1$.  \label{fig:hkcf}}
\end{figure}

Fig.~\ref{fig:hkghf}-(b) shows the improvement of GHF versus CF. As shown in
this figure, GHF outperforms CF in a triangular region for values of
$\alpha>1/2$ for a symmetric interference channel. CF is specially not suited at
$\alpha=2/3$. Notice from \eqref{eq:cf1sumrate} and \eqref{eq:cf2sumrate} that
for $\rho\leq 1/3$, CF gives zero improvement in the sum rate asymptotically; see Fig.~\ref{fig:hkcf}-(a). It becomes further
clear as to why $\alpha=2/3$ is special when we compare the two decoding
orders for CF.

Fig.~\ref{fig:hkcf}-(a) and Fig.~\ref{fig:hkcf}-(b) compare the asymptotic sum rate improvement with CF for
different decoding orders. When the relay observation  is reconstructed first, CF gives
zero gain for $1/2<\alpha<2/3$ and $\rho<2\alpha-1$. But we can recover from
this zero-gain regime if we switch the decoding order as shown in
Fig.~\ref{fig:hkcf}-(b) for CF with optimal decoding order. Notice that only for
$1/2<\alpha<2/3$ we need to switch the decoding order, and thus, $\alpha=2/3$
remains as a transition point; for $\alpha>2/3$, the optimal decoding order is
to reconstruct the relay observation first, and for $\alpha<2/3$, the optimal order is to decode the intended common message first. This leaves no successive decoding option
 at $\alpha=2/3$ to benefit from the relay.

\subsubsection{An Asymmetric Example}
When the constraints on $R_0$ in \eqref{eq:r0cond} are not satisfied, the GHF strategy of Theorem~\ref{thm:improvement} is not optimal in general.  The following example shows that in certain regimes, a CF strategy with a different quantization parameter outperforms the GHF strategy of Theorem~\ref{thm:improvement}. This is because the quantization parameter in GHF strategy of Theorem~\ref{thm:improvement} is not optimized over all values of $q$ to maximize the achievable rate, but it is chosen to satisfy a constant gap result for the entire capacity region, when $R_0$ satisfies $\eqref{eq:r0cond}$.  However, note that if $q$ is chosen to be the one used in CF, then GHF also achieves the same rate as CF (since the key difference between CF and GHF is that GHF uses joint decoding, whereas CF uses successive decoding).


Consider the asymptotic scenario with $\SNR\ra \infty$ for fixed $\alpha_1,\alpha_2$ and  $\beta_1,\beta_2\ra 1^-$ for an asymmetric interference channel, and assume that $\alpha_1<\rho<\alpha_2$, and $\alpha_1+\alpha_2<1$. In this case, from \eqref{eq:app:ghf} and \eqref{eq:app:cf1}, the asymptotic achievable sum rate for GHF and CF with relay codeword decoded first are given as:
\begin{align}
R_1+R_2&\leq \log\SNR\NoN\\
R_1+R_2&\leq \frac{1}{2}(2-\alpha_2)\log\SNR+\frac{1}{2}\alpha_1\log\SNR\NoN\\
R_1+R_2&\leq \frac{1}{2}(2-\alpha_1-\alpha_2)\log\SNR+\alpha_1\log\SNR,
\intertext{and,}
R_1+R_2&\leq \log\SNR\NoN\\
R_1+R_2&\leq \frac{1}{2}(2-\alpha_2)\log\SNR+\frac{1}{2}\rho\log\SNR\NoN\\
R_1+R_2&\leq \frac{1}{2}(2-\alpha_1-\alpha_2)\log\SNR+\frac{1}{2}\rho\log\SNR+\frac{1}{2}\alpha_1\log\SNR,
\end{align}
respectively.

 In this case, CF with decoding the relay quantized codeword first outperforms GHF strategy with $q$ given in \eqref{eq:qweak}, when $\rho>\alpha_1$. Notice, however, that we could have used the same $q$ used in CF for GHF. In fact, we could optimized $q$ in GHF.  As an alternative to optimizing $q$, one may also choose $q$ among the following strategies
\begin{align}
  q&=\frac{1}{2^{2R_0}-1}\min\bigl\{\var(Y_r\vert Y_1W_1),\var(Y_r\vert
  Y_2W_2)\bigr\}\NoN\\
    q&=\frac{1}{2^{2R_0}-1}\max\bigl\{\var(Y_r\vert Y_1W_1),\var(Y_r\vert
  Y_2W_2)\bigr\}\NoN\\
  q&=\frac{1}{2^{2R_0}-1}\min\bigl\{\var(Y_r|Y_1),\var(Y_r|Y_2)\bigr\}\NoN\\
  q&=\frac{1}{2^{2R_0}-1}\max\bigl\{\var(Y_r|Y_1),\var(Y_r|Y_2)\bigr\},
\end{align}
obtained by considering quantization strategy in CF with all possible decoding orders, with an added flexibility of using $\min$ instead of $\max$, which becomes possible via joint decoding in GHF.


\clearpage




\section{Concluding Remarks \label{sec:conc}}
We studied the two-user Gaussian interference channel with an out-of-band
relay forwarding a common message of a limited rate over a noiseless link to
the two destinations. We focused on oblivious relay strategies where the encoding strategy at source nodes  is independent of the relay presence (apart from the rate allocation which is higher in relay presence). For relay rates below certain threshold, the entire capacity region of this channel was characterized to within a constant gap. In this regime, a carefully designed quantize-and-forward strategy can be very efficient, in the sense that  every bit relayed improves the sum rate by close to two bits. 

The interference channel with a relay is different from the classic single-user
relay channel studied in \cite{cover_elgamal} in that the relay simultaneously
serves more than one destination node. In such scenarios, we showed that
conventional  source-coding with side information is inefficient in general for
relay quantization. We employed an extended class of quantize-and-forward
strategies and introduced a list decoding strategy which emphasizes on decoding
the source message. This
approach was compared in details with the conventional compress-and-forward
with successive decoding with an optimal decoding order. In particular, we found that even with optimal 
decoding order, conventional CF with successive decoding achieves similar gains like joint decoding in certain regimes, in particular in regimes where a two-for-one gain is attainable. However, it was also shown that successive decoding results in unbounded gaps to capacity, for example, in
symmetric interference channel with $\log\INR/\log\SNR=2/3$.

The constant-gap results in this paper are valid when the rate of the relay link is below a threshold. Intuitively, we expect that as the rate of relay link increases, the interference channel behaves more like a SIMO interference channel with two antennas at each destination, since the relay can more accurately communicate its observation to the two destinations. Further, with a link of a higher rate, the relay can split its excess rate and forward dedicated messages for each user. In this case, we may also need to modify the power splitting strategy at the source nodes. However, we focused in this paper on relay strategies where the source nodes are oblivious to the relay. We observed that there are many operating regimes even for a limited rate relay link.  For larger relay link rates, characterization of the capacity region  is a more complicated task and is left for future work.

\appendices
\section{Proof of Theorem~\ref{thm:ghf}\label{app:proofghf}}
The source transmits $nR(B-1)$ bits over $B$ blocks of $n$ symbols each. In the last block no bits are transmitted. As $B\rightarrow \infty$, $nR(B-1)$ divided by the number of symbols $nB$ tends to $R$.

{\it Codebook Generation:}
Randomly and independently generate $2^{nR}$ codewords $X^n(w)$ of
length $n$ indexed by $ w\in\{1,\ldots,2^{nR}\}$ according to
$\prod_{i=1}^n p(x_i)$. Fix a $p(\hat{y}_r|y_r)$ such 
Randomly and independently generate $2^{n(I(Y_r;\hat{Y_r})+\epsilon)}$
codewords $\hat{Y_r}^n(r)$, $r\in\{1,\ldots,2^{n(I(Y_r;\hat{Y_r})+\epsilon)}\}$ of length $n$ according to $\prod_{i=1}^n p(u_i)$. We shall also need a
random partition of the $\hat{Y_r}^n$ codewords into bins. Randomly partition
the set $\{1,2,\ldots,2^{n(I(Y_r:\hat{Y_r})+\epsilon)}\}$ into $2^{nR_0}$ bins
$\B_l,l\in\{1,\ldots,2^{nR_0}\}$ each of size $2^{n(I(Y_r:\hat{Y_r})-R_0+\epsilon)}$.

{\it Encoding:}
In block $i$, the source sends $X^n(w_i)$.
Having observed $Y_r^n(i-1)$ in block $i-1$, the relay finds a
codeword $\hat{Y_r}^n(t_i)$, $t_i\in\{1,\ldots,2^{n(I(Y_r;\hat{Y_r})+\epsilon)}\}$,
such that $(\hat{Y_r}^n(t_i),Y_r^n(i-1))$ is $\epsilon$-strongly typical (see
\cite[Section~13.6]{cover_elements} for definition of strong
typicality). The relay sends $k$, the bin index of $t_i$ over the digital
channel to the destination in block $i$, (i.e.\ $t_i \in \B_{k}$).

{\it Decoding:}
In block $i$, the destination decodes the source message of block
$i-1$ in following steps:
\begin{enumerate}
\item Upon receiving $k$, the destination forms an index list $\Li$
of possible $\hat{Y_r}^n$-codewords by identifying indices $r\in \B_{k}$
such that $(\hat{Y_r}^n(r), Y^n(i-1)$ are $\epsilon$-strongly typical.
\item Destination finds a source codeword that is consistent with its
own observation $Y^n(i-1)$ and $\Li$ by finding $\hat{w} \in
\{1,\ldots,2^{nR}\}$ such that the three-tuple
$(X^n(\hat{w}),\hat{Y_r}^n(m),Y^n(i-1))$ is $\epsilon$-strongly typical for
some $m\in \Li$.
\end{enumerate}

{\it Analysis of Probability of Error:} By the well-known random coding symmetrization argument \cite{cover_elements}, we can
assume that $X^n(1)$ is sent over all blocks.  Since decoding events
in different blocks are independent, we can also focus on block $i$ to
analyze probability of error, and drop the time indices. The error
events are as follows:

\begin{enumerate}
\item[$E_1$:] $(X^n(1),Y^n_r,Y^n)\notin\EPS$,
\item[$E_2$:] $\nexists t\in\{1,\ldots,2^{n(I(\hat{Y_r};Y_r)+\epsilon}\}$ such
	that $(\hat{Y_r}^n(t),Y_r^n)\in\EPS$,
\item[$E_3$:] $\nexists s\in \Li$ such that
	$\bigl(X^n(1),Y^n,\hat{Y_r}^n(s)\bigr)\in\EPS$.
\item[$E_4$:] $\exists m,w': m\in \Li,w'\in\{1,\ldots,2^{R}\}$, $w'\neq 1$,
	such that $(X^n(w'),\hat{Y_r}^n(m),Y^n)\in\EPS$,
\end{enumerate}
where $\EPS$ denotes the set of $\epsilon$-strongly typical sequences of length $n$ for a given joint probability \cite{cover_elements}.

For $n$ sufficiently large, $P(E_1)\leq \epsilon$  for arbitrarily
small $\epsilon>0$ \cite[Lemma 10.6.1]{cover_elements}. Following the
argument of \cite[Section 10.6]{cover_elements},  $P(E_2\cap
E_1^c)\leq \epsilon$ for sufficiently large $n$, since the number of
$\hat{Y_r}^n$ codewords is more than $2^{nI(\hat{Y_r};Y_r)}$.


By the the Markov Lemma \cite[Lemma 15.8.1]{cover_elements},  since $(X,Y)-Y_r-\hat{Y_r}$ forms
a Markov chain, we have $(X^n,Y^n,\hat{Y}_r^n)\in\EPS$ for $(X^n,Y^n,Y_r^n)\in\EPS$, i.e., $P(E_3\cap\bigcap_{j=1}^2E_j)<\epsilon$ for sufficiently large $n$.

To bound the probability of $E_4$, note that for $X^n(w')$ drawn i.i.d.
$\sim\prod p(x_i)$ and independent of $\epsilon$-strongly typical pair
$(\hat{Y_r}^n(m),Y^n)$, the probability that $(X^n(w'),\hat{Y_r}^n(m),Y^n) \in \EPS$ is
less than $2^{-n(I(X;Y,\hat{Y_r})-\epsilon)}$ for sufficiently large $n$ and
arbitrarily $\epsilon>0$ \cite[Lemma 10.6.2]{cover_elements}.  Let $A$
be the event that $(X^n(w'),\hat{Y_r}^n(m),Y^n) \in\EPS$ for some $m\in\Li$
and $w'\in\{1,\ldots,2^{nR}\},w'\neq 1$, assuming that $E_i$ does not
occur for $i=1,\cdots,3$. We have
\begin{align}
P\left(\bigcap_{j=1}^4 E_j\right)&=P(A)\NoN\\
&=\sum_lP\Bigl(A\Big|\lvert{\Li}\rvert=l\Bigr)P\Bigl(\lvert{\Li}\rvert=l\Bigr)\NoN\\
&\leq \sum_lP\bigl(\lvert\Li\rvert=l\bigr)\sum_{m\in\Li,w'}2^{-n(I(X;Y,\hat{Y_r})-\epsilon)}\NoN\\
&= \sum_lP\bigl(\lvert\Li\rvert=l\bigr)\cdot l\cdot 2^{nR}\cdot 2^{-n(I(X;Y,\hat{Y_r})-\epsilon)} \NoN\\
&= 2^{nR}2^{-n(I(X;Y,\hat{Y_r})-\epsilon)}\Ex\lvert\Li\rvert,\label{eq:bound-x}
\end{align}
where $\lvert\Li\rvert$ represents the cardinality of $\Li$.

Now, the method employed in \cite[Lemma 3]{cover_elgamal} can be used
to find an upper bound on  $\Ex\lvert{\Li}\rvert$. Recall that $\Li$ is the
list of $\hat{Y_r}^n(r)$ codewords with $r\in\B_k$ and  $(\hat{Y_r}^n,Y)$
$\epsilon$-strongly typical. Let
\begin{align}
\psi(r|Y^n)=\left\{\begin{array}{ll}
    1&(\hat{Y_r}^n(r),Y^n) \text{ is $\epsilon$-strongly typical,}\NoN\\
0&\text{otherwise.}\NoN\end{array}\right.
\end{align}
Then, $\lvert\Li\rvert$ can be expressed as:
\begin{equation}
\lvert{\Li}\rvert=\sum_{r\in\B_k}\psi(r|Y^n).
\end{equation}
We have
\begin{align}
\Ex\lvert{\Li}\rvert&=\Ex\psi(t|Y^n)+\sum_{r\neq t, r\in\B_k}\Ex\psi(r|Y^n)\NoN\\
&=P\Bigl(\psi(t|Y^n)=1\Bigl)+\sum_{r\neq t,r\in\B_k}P\Bigr(\psi(r|Y^n)=1\Bigl)\NoN\\
&\overset{(*)}\leq 1+(2^{\lvert{\B_k}\rvert}-1)2^{-n(I(\hat{Y_r};Y)-\gamma)} \nonumber\\
&\leq 1+2^{n(I(\hat{Y_r};Y_r)-R_0-I(\hat{Y_r};Y)+\epsilon+\gamma)}\NoN\\
&=1+2^{n(I(\hat{Y_r};Y_r|Y)-R_0+\epsilon+\gamma)},\label{eq:size-L}
\end{align}
where $\lvert\B_k\rvert$ denotes the cardinality of $\B_k$, and $(*)$ follows from \cite[Lemma 10.6.2]{cover_elements} for
sufficiently large $n$ and arbitrarily small $\gamma>0$.

From \eqref{eq:size-L} and \eqref{eq:bound-x}, we have:
\begin{align}
P(A)&<2^{nR}\cdot 2^{-n(I(X;Y,\hat{Y}_r)-\epsilon)}\left(1+2^{n(I(\hat{Y_r};Y_r|Y)-R_0+\epsilon+\gamma)}\right)\NoN\\
&=2^{n(R-I(X;Y,\hat{Y}_r)+\epsilon)}+2^{n(R-I(X;Y,\hat{Y}_r)+ I(\hat{Y_r};Y_r|Y)-R_0+\epsilon+\gamma)},
\end{align}
which tends to zero asymptotically for large $n$ provided that:
\begin{align}
R&<I(X;Y,\hat{Y_r}) \label{eq:1}\\
R&<I(X;Y,\hat{Y_r}) +R_0-I(\hat{Y_r};Y_r|Y)\label{eq:18}.
\end{align}
Note \eqref{eq:18} can be simplified
as follows
\begin{align}
 R & < I(X;Y)+I(X;\hat{Y_r}|Y)+R_0-I(\hat{Y_r};Y_r|Y)\NoN\\
   & \overset{(a)}=I(X;Y)+I(X;\hat{Y_r}|Y)+R_0-h(\hat{Y_r}|Y)+h(\hat{Y_r}|Y_r)\NoN\\
   & \overset{(b)}=I(X;Y)+R_0-\bigl(h(\hat{Y_r}|X,Y)-h(\hat{Y_r}|Y_r,X,Y)\bigr)\NoN\\
   & = I(X;Y)+R_0-I(\hat{Y_r};Y_r|X,Y)\NoN,
\end{align}
where (a) and (b) follow from the Markov chain $\hat{Y_r}-Y_r-(X,Y)$.
This proves the theorem.

\section{Proof of Theorem~\ref{thm:5}\label{sec:proof}}
To prove the achievability, consider a memoryless interference channel defined by $p(y_1,y_2|x_1,x_2)$, where the two users employ  the HK strategy developed in \cite{chong_motani}. In this strategy,  the first source uses an auxiliary random variable $W_1$ to randomly generate $2^{nT_1}$ cloud center codewords $W_1^n(j),j=1,\ldots,2^{nT_1}$ of length $n$ according to $p(w_1|q)$ where $Q\sim p(q)$ represents a time-sharing auxiliary random variable. For each $W_1^n(j)$, user one generates $2^{nS_1}$ codewords $X_1^n(j,k),k=1,\ldots,2^{nS_1}$ of length $n$ according to $p(x_1|w_1,q)$. Similarly, user two generates $2^{nT_2}$ cloud center codewords $W_2^n(l),l=1,\ldots,2^{nT2}$ according to $p(w_2|q)$, each surrounded by $2^{nS_2}$ random codewords $X_2^n(l,m)$ generated according to $p(x_2|w_2,q)$. In \cite{chong_motani}, it is shown that a rate pair $(R_1,R_2)=(S_1+T_1,S_2+T_2)$ is achievable provided that (see \cite[(167)-(178)]{kobayashi_han_extension})
\begin{subequations}\label{eq:HK}
  \iftwocol
\begin{align}
S_1&\leq a_1,\label{eq:s1}\\
S_1+T_1&\leq d_1,\label{eq:s1t1}\\
S_1+T_2&\leq e_1,\\
S_1+T_1+T_2&\leq g_1,\\
-S_1&\leq 0,\\
-T_1&\leq 0\\
\intertext{and}
S_2&\leq a_2\\
S_2+T_2&\leq d_2\\
S_2+T_1&\leq e_2\\
S_2+T_1+T_2&\leq g_2\\
-S_2&\leq 0\\
-T_2&\leq 0.
\end{align}
\else
\begin{align}
S_1&\leq a_1,\label{eq:s1} & S_2&\leq a_2,\\
S_1+T_1&\leq d_1,\label{eq:s1t1} & S_2+T_2&\leq d_2,\\
S_1+T_2&\leq e_1, & S_2+T_1&\leq e_2,\\
S_1+T_1+T_2&\leq g_1, & S_2+T_1+T_2&\leq g_2,\\
-S_1&\leq 0, & -S_2&\leq 0,\\
-T_1&\leq 0, & -T_2&\leq 0.
\end{align}
\fi
\end{subequations}
In the GHF strategy, the relay quantizes its observation $Y_r$ using the auxiliary random variable $\hat{Y_r}$ and sends a bin index of rate $R_0$ for the quantized relay codeword to both destinations. The bin index of $\hat{Y_r}$ improves the achievable rates for $S_i$ and $T_i$, $i=1,2$, in \eqref{eq:HK} and consequently the achievable rate of each user.

Using Theorem \ref{thm:2},  we can find the new constraints for $S_i,T_i,i=1,2$ when GHF is used. Assume without loss of generality that $X_1^n(1,1)$ and $X_2^n(1,1)$ are sent by the two sources. Note that, for example, the first constraint on $S_1$ in \eqref{eq:s1} corresponds to an error event $\A$ where the first user decodes a wrong private message of rate $S_1$ while the common messages (encoded by $W_1^n(1)$ and $W_2^n(1)$) are decoded correctly. A conditional version of Theorem~\ref{thm:2} for given $W_1$ and $W_2$ guarantees that with the help of the bin index sent for $\hat{Y_r}$ from the relay, the probability of the event $\A$ vanishes asymptotically provided that $S_1$ satisfies
\iftwocol
\begin{align}
S_1&\leq a_1+\Delta a_1-\Delta_1,
\intertext{where:}
\Delta a_1&=\min\bigl\{R_0,I(\hat{Y_r};Y_r|Y_1,W_1,W_2,Q)\bigr\},
\intertext{and}
\Delta_1&=\min\bigl\{R_0,I(\hat{Y_r};Y_r|Y_1,X_1,W_1,W_2,Q)\bigr\}\NoN\\
&\overset{(a)}=\min\bigl\{R_0,I(\hat{Y_r};Y_r|Y_1,X_1,W_2,Q)\bigr\},\label{eq:delta}
\end{align}
\else
\begin{align}
S_1&\leq a_1+\Delta a_1-\Delta_1,
\intertext{where:}
\Delta_1&=\min\bigl\{R_0,I(\hat{Y_r};Y_r|Y_1,X_1,W_1,W_2,Q)\bigr\} \overset{(a)}=\min\bigl\{R_0,I(\hat{Y_r};Y_r|Y_1,X_1,W_2,Q)\bigr\},\label{eq:delta}
\end{align}
\fi
where (a) follows from the Markov chain $W_1-(X_1,Q)-(W_2,Y_1,Y_r,\hat{Y_r})$.

Similarly, \eqref{eq:s1t1} corresponds to an event $\B$ where both common and private messages of rates $S_1$ and $T_1$ are decoded incorrectly by user one, while the common message of user two (encoded by $W_2^n(1)$) is decoded correctly. Again, a conditional version of Theorem \ref{thm:2} for given $W_2$ ensures that the probability of the event $\B$ vanishes asymptotically provided that
\begin{align}
S_1+T_1\leq d_1+\Delta d_1-\Delta_1.
\end{align}

Using similar arguments for other constraints in \eqref{eq:HK}, we find the following achievable rate region for an interference channel with a digital relay:
\begin{subequations}\label{eq:relay_HK}
\iftwocol
\begin{align}
S_1&\leq a_1+\Delta a_1-\Delta_1,\label{eq:s1}\\
S_1+T_1&\leq d_1 + \Delta d_1-\Delta_1,\label{eq:s1t1}\\
S_1+T_2&\leq e_1 + \Delta e_1-\Delta_1,\\
S_1+T_1+T_2&\leq g_1 + \Delta g_1-\Delta_1,\\
-S_1&\leq 0,\\
-T_1&\leq 0\\
\intertext{and}
S_2&\leq a_2 + \Delta a_2-\Delta_2\\
S_2+T_2&\leq d_2 + \Delta d_2-\Delta_2\\
S_2+T_1&\leq e_2 + \Delta e_2-\Delta_2\\
S_2+T_1+T_2&\leq g_2 + \Delta g_2-\Delta_2\\
-S_2&\leq 0\\
-T_2&\leq 0
\end{align}
\else
\begin{align}
S_1&\leq a_1+\Delta a_1-\Delta_1,\label{eq:s1}&S_2&\leq a_2 + \Delta a_2-\Delta_2,\\
S_1+T_1&\leq d_1 + \Delta d_1-\Delta_1,\label{eq:s1t1}& S_2+T_2&\leq d_2 + \Delta d_2-\Delta_2,\\
S_1+T_2&\leq e_1 + \Delta e_1-\Delta_1,&S_2+T_1&\leq e_2 + \Delta e_2-\Delta_2,\\
S_1+T_1+T_2&\leq g_1 + \Delta g_1-\Delta_1,&S_2+T_1+T_2&\leq g_2 + \Delta g_2-\Delta_2,\\
-S_1&\leq 0,&-S_2&\leq 0,\\
-T_1&\leq 0,& -T_2&\leq 0.
\end{align}
\fi
\end{subequations}

The above region can be further simplified using Fourier-Motzkin algorithm \cite{kobayashi_han_extension}. First note that we have:
\begin{align}
d_i&\leq g_i,\label{eq:dg}\\
a_i&\leq e_i\leq g_i,
\end{align}
for $i=1,2$. Next, we also have:
\begin{align}\
\Delta d_i&\leq \Delta g_i,\\
\Delta a_i&\leq \Delta e_i\leq \Delta g_i\label{eq:dddg},
\end{align}
since, for example,
\iftwocol
\begin{align}
\Delta a_1&=I(\hat{Y_r};Y_r|Y_1,W_1,W_2,Q)\NoN\\
&\overset{(a)}=H(\hat{Y_r}|Y_1,W_1,W_2,Q)-H(\hat{Y_r}|Y_r,Q)\NoN\\
&\leq H(\hat{Y_r}|Y_1,W_2,Q)-H(\hat{Y_r}|Y_r,Q)\NoN\\
&\overset{(a)}=I(\hat{Y_r};Y_r|Y_1,W_1,Q)=\Delta e_1\NoN\\
&\leq H(\hat{Y_r}|Y_1,Q)-H(Y|Y_r,Q)=\Delta g_1\NoN
\end{align}
\else
\begin{align}
\Delta a_1&=I(\hat{Y_r};Y_r|Y_1,W_1,W_2,Q)\overset{(a)}=H(\hat{Y_r}|Y_1,W_1,W_2,Q)-H(\hat{Y_r}|Y_r,Q)\NoN\\
&\leq H(\hat{Y_r}|Y_1,W_2,Q)-H(\hat{Y_r}|Y_r,Q)\NoN\\
&\overset{(a)}=I(\hat{Y_r};Y_r|Y_1,W_1,Q)=\Delta e_1\NoN\\
&\leq H(\hat{Y_r}|Y_1,Q)-H(Y|Y_r,Q)=\Delta g_1,\NoN
\end{align}
\fi
where (a) follows from the Markov chain $\hat{Y_r}-(Y_r,Q)-(Y_1,W_1,W_2)$.

Now, by following exactly the same steps in \cite[Section~III]{kobayashi_han_extension}, with $a_i,d_i,e_i,g_i$ replaced by $a_i+\Delta a_i-\Delta_i,d_i+\Delta d_i -\Delta_i,e_i+\Delta e_i-\Delta_i,g_i+\Delta g_i-\Delta_i$, respectively,  we get the following achievable rate for $(R_1,R_2)$ from \eqref{eq:relay_HK} by using Fourier-Moztkin elimination:
\begin{subequations}\label{eq:relay_HK2}
\begin{align}
R_1&\leq d_1+\Delta d_1 - \Delta_1\label{eq:d1}\\
R_1&\leq a_1+\Delta a_1- \Delta_1  +e_2+\Delta e_2  -\Delta_2\label{eq:ext1}\\
R_2&\leq d_2 +\Delta d_2 -\Delta_2\label{eq:d2}\\
R_2&\leq e_1+\Delta e_1-\Delta_1+a_2+\Delta a_2 -\Delta_2 \label{eq:ext2}\\
R_1+R_2&\leq a_1+\Delta a_1- \Delta_1+g_2+\Delta g_2-\Delta_2\\
R_1+R_2&\leq g_1+\Delta g_1-\Delta_1+a_2+\Delta a_2- \Delta_2\\
R_1+R_2&\leq e_1+\Delta e_1- \Delta_1+e_2+\Delta e_2-\Delta_2\\
2R_1+R_2&\leq a_1+\Delta a_1 + g_1+\Delta g_1 -2\Delta_1 +  e_2+\Delta e_2  -\Delta_2\\
R_1+2R_2&\leq  e_1+\Delta e_1 - \Delta_1+a_2+\Delta a_2 +  g_2+\Delta g_2 -2\Delta_2 \\
R_1&>0\\
R_2&>0,
\end{align}
\end{subequations}
for some $(Q,W_1,W_2,X_1,X_2,\hat{Y_r},Y_r)\sim p(q)p(x_1,w_1|q)p(x_2,w_2|q)p(\hat{y}_r|y_r,q)$.

We can further simplify the above region by noting that \eqref{eq:ext1} and \eqref{eq:ext2} can be eliminated through time sharing between three rate-splitting strategies. Let $p(q,w_1,w_2,x_1,x_2)$ denote a particular distribution for $W_1,W_2,X_1,X_2,Q$. Construct new distributions $p^*$ and $p^{**}$ from $p$ by eliminating $w_1$ and $w_2$, respectively, as:
\begin{subequations}\label{eq:time-share}
\begin{align}
p^*(q,w_1,w_2,x_1,x_2)=\sum_{w_1}p(q)p(x_1,w_1)p(x_2,w_2).\\
P^{**}(q,w_1,w_2,x_1,x_2)=\sum_{w_2}p(q)p(x_1,w_1)p(x_2,w_2).
\end{align}
\end{subequations}
By \eqref{eq:relay_HK_simple},  the rate pair $(R_1,R_2)$ satisfying \eqref{eq:relay_HK_simple} is achievable using an HK strategy along with GHF for an input distribution $p(q,w_1,w_2,x_1,x_2)$ provided that \eqref{eq:ext1} and \eqref{eq:ext2} are also satisfied.

If \eqref{eq:ext1} is not satisfied, then $(R_1,R_2)$ can be achieved using the input distribution $p^*(q,w_1,w_2,x_1,x_2)$ obtained from $p$ according to \eqref{eq:time-share}. By setting $W_1=\phi$ in ~\eqref{eq:relay_HK_simple}, all rate pairs $(R_1,R_2)$ satisfying the following constraints are achievable using $p^*$:
\iftwocol
\begin{align}
 R_1&\leq d_1+\Delta d_1-\Delta_1\label{eq:11}\\
 R_1&\leq d_1+\Delta d_1-\Delta_1+I(Y_2;X_2|W_2,Q)+\NoN\\
&\qquad \Delta e_2-\Delta'_2\label{eq:22}\\
 R_2&\leq I(X_2;Y_2|Q)+\Delta g_2-\Delta'_2\label{eq:33}\\
 R_2&\leq g_1+\Delta g_1-\Delta_1+I(X_2;Y_2|W_2,Q)+\NoN\\
&\qquad \Delta e_2-\Delta'_2\label{eq:44}\\
 R_1+R_2&\leq d_1+\Delta d_1-\Delta_1+I(X_2:Y_2|Q)+\NoN\\
&\qquad \Delta g_2-\Delta'_2\label{eq:55}\\
R_1+R_2&\leq g_1+\Delta g_1-\Delta_1+I(X_2;Y_2|W_2,Q)+\NoN\\
&\qquad \Delta e_2-\Delta'_2\label{eq:66}\\
R_1+R_2&\leq g_1+\Delta g_1-\Delta_1+I(X_2;Y_2|W_2,Q)+\NoN\\
&\qquad \Delta e_2-\Delta'_2\label{eq:77}\\
2R_1+R_2&\leq d_1+\Delta d_1+g_1+\Delta g_1-2\Delta_1\NoN\\
&\quad+I(X_2;Y_2|W_2,Q)+\Delta e_2-\Delta'_2\label{eq:88}\\
R_1+2R_2&\leq g_1+\Delta g_1-\Delta_1+I(X_2;Y_2|W_2,Q)+\Delta e_2\NoN\\
&\quad+I(X_2;Y_2|Q)+\Delta g_2-2\Delta'_2\label{eq:99},
\end{align}
\else
\begin{align}
 R_1&\leq d_1+\Delta d_1-\Delta_1\label{eq:11}\\
 R_1&\leq d_1+\Delta d_1-\Delta_1+I(Y_2;X_2|W_2,Q)+\Delta e_2-\Delta'_2\label{eq:22}\\
 R_2&\leq I(X_2;Y_2|Q)+\Delta g_2-\Delta'_2\label{eq:33}\\
 R_2&\leq g_1+\Delta g_1-\Delta_1+I(X_2;Y_2|W_2,Q)+\Delta e_2-\Delta'_2\label{eq:44}\\
 R_1+R_2&\leq d_1+\Delta d_1-\Delta_1+I(X_2:Y_2|Q)+\Delta g_2-\Delta'_2\label{eq:55}\\
R_1+R_2&\leq g_1+\Delta g_1-\Delta_1+I(X_2;Y_2|W_2,Q)+\Delta e_2-\Delta'_2\label{eq:66}\\
R_1+R_2&\leq g_1+\Delta g_1-\Delta_1+I(X_2;Y_2|W_2,Q)+\Delta e_2-\Delta'_2\label{eq:77}\\
2R_1+R_2&\leq d_1+\Delta d_1+g_1+\Delta g_1-2\Delta_1+I(X_2;Y_2|W_2,Q)+\Delta e_2-\Delta'_2\label{eq:88}\\
R_1+2R_2&\leq g_1+\Delta g_1-\Delta_1+I(X_2;Y_2|W_2,Q)+\Delta e_2+I(X_2;Y_2|Q)+\Delta g_2-2\Delta'_2\label{eq:99},
\end{align}
\fi
where $\Delta_2'=\min\{R_0,I(\hat{Y_r};Y_r|X_2,Y_2,Q)\}$. The above region can be simplified by removing redundant constraints. Thus, \eqref{eq:22} is redundant due to \eqref{eq:11}. Next, \eqref{eq:77} is redundant due to \eqref{eq:66}, and \eqref{eq:55} is redundant due to \eqref{eq:11} and \eqref{eq:33}.  Also, \eqref{eq:88} is redundant due to \eqref{eq:11} and \eqref{eq:66}. Finally, \eqref{eq:99} is redundant due to \eqref{eq:33} and \eqref{eq:77}, and \eqref{eq:44} is redundant due to \eqref{eq:66}. In summary, the following region is achievable using $p^*$:
\begin{subequations}\label{eq:p*}
\begin{align}
R_1&\leq d_1+\Delta d_1-\Delta_1\label{eq:s1}\\
R_2&\leq I(X_2;Y_2|Q)+\Delta g_2-\Delta'_2\label{eq:s2}\\
R_1+R_2&\leq g_1+\Delta g_1-\Delta_1+I(X_2;Y_2|W_2,Q)+ \Delta e_2-\Delta'_2\label{eq:s4}.
\end{align}
\end{subequations}
Now, we can prove that if $(R_1,R_2)$ satisfies \eqref{eq:relay_HK_simple} while \eqref{eq:ext1} is violated, $(R_1,R_2)$ satisfies \eqref{eq:p*} and hence  is  achievable by using input distribution $p^*$. If \eqref{eq:ext1} is violated, we have:
\begin{align}
-R_1&\leq -a_1-\Delta a_1+\Delta_1-e_2-\Delta e_2+\Delta_2\label{eq:viol}.
\end{align}
Now, \eqref{eq:s1} follows from \eqref{eq:r1}. From \eqref{eq:viol} and \eqref{eq:r3}, we have:
\iftwocol
\begin{align}
R_2&\leq I(Y_1;X_1|W_1,W_2,Q)+I(Y_2;X_2,W_1,Q)\NoN\\
&\quad -I(Y_1;X_1|W_1,W_2,Q)-I(Y_2;X_2W_1|W_2)\NoN\\
&\quad +\Delta a_1+\Delta g_2 -\Delta e_2-\Delta a_1\NoN\\
&=I(Y_2;X_2|Q)+I(Y_2;W_1|X_2,Q)\NoN\\
&\quad -I(Y_2;X_2|W_2,Q)-I(Y_2;W_1|X_2,W_2,Q)\NoN\\
&\quad +\Delta g_2 -\Delta e_2\NoN\\
&\overset{(a)}=I(Y_2;X_2|Q)-I(Y_2;X_2|W_2,Q)\NoN\\
&\quad +\Delta g_2 -\Delta e_2\NoN\\
&\leq I(Y_2;X_2|Q)+\Delta g_2 -\Delta e_2\NoN\\
&\overset{(b)}\leq I(Y_2;X_2|Q)+\Delta g_2 - \Delta_2'\NoN,
\end{align}
\else
\begin{align}
R_2&\leq I(Y_1;X_1|W_1,W_2,Q)+I(Y_2;X_2,W_1,Q) -I(Y_1;X_1|W_1,W_2,Q)-I(Y_2;X_2W_1|W_2)\NoN\\
&\hspace{0.5\textwidth}  +\Delta a_1+\Delta g_2 -\Delta e_2-\Delta a_1\NoN\\
&=I(Y_2;X_2|Q)+I(Y_2;W_1|X_2,Q) -I(Y_2;X_2|W_2,Q)-I(Y_2;W_1|X_2,W_2,Q)+\Delta g_2 -\Delta e_2\NoN\\
&\overset{(a)}=I(Y_2;X_2|Q)-I(Y_2;X_2|W_2,Q) +\Delta g_2 -\Delta e_2\NoN\\
&\leq I(Y_2;X_2|Q)+\Delta g_2 -\Delta e_2\NoN\\
&\overset{(b)}\leq I(Y_2;X_2|Q)+\Delta g_2 - \Delta_2'\NoN,
\end{align}
\fi
where (a) follows since $I(Y_2;W_1|X_2,W_2,Q)=I(Y_2;W_1|X_2,Q)$ from the Markov chain $W_2-(X_2,Q)-(W_1,Y_2)$, and (b) follows since $\Delta e_2-\Delta_2'\geq 0$. Thus, \eqref{eq:s2} follows from \eqref{eq:viol} and \eqref{eq:r3}.

Finally, \eqref{eq:s4} follows from \eqref{eq:viol} and \eqref{eq:r6}, since we have:
\begin{align}
R_1+R_2&\leq a_1+\Delta a_1+g_1+\Delta g_1-2\Delta_1+e_2+\Delta e_2\NoN\\
&\quad -\Delta_2-a_1-\Delta a_1+\Delta_1-e_2-\Delta e_2+\Delta_2 \NoN\\
&=g_1+\Delta g_1-\Delta_1\NoN\\
&\leq g_1+\Delta g_1-\Delta_1+I(X_2;Y_2|W_2,Q)+\Delta e_2-\Delta_2'\NoN,
\end{align}
which completes the proof of achievability of $(R_1,R_2)$ under $p^*$ if \eqref{eq:ext1} is violated. By symmetry, $(R_1,R_2)$ is again achievable under $p^{**}$ if \eqref{eq:ext2} is violated. Thus, any rate pair $(R_1,R_2)$ satisfying \eqref{eq:relay_HK_simple} for some $p$ is achievable under input distribution $p$, or $p^*$, or $p^{**}$. This completes the proof of Theorem~\ref{thm:5}.


\section{Upper Bounds  \label{sec:bounds}}

\begin{theorem}\label{thm:boundsweak}
In the weak interference regime where $\INR_1<\SNR_1$, $\INR_2<\SNR_2$,  the capacity region of the Gaussian interference channel with an out-of-band relay  of rate $R_0$ is contained in the following region:
\begin{align}
R_1&\leq \frac{1}{2}\log\left(1+\SNR_1\right)+K_1\NoN\\
R_2&\leq \frac{1}{2}\log\left(1+\SNR_2\right)+K_2\NoN\\
R_1+R_2&\leq \frac{1}{2}\log\left(1+\SNR_1\right)+\frac{1}{2}\log\left(1+\frac{\SNR_2}{1+\INR_2}\right)+R_0+K_1\NoN\\
R_1+R_2&\leq \frac{1}{2}\log\left(1+\SNR_2\right)+\frac{1}{2}\log\left(1+\frac{\SNR_1}{1+\INR_1}\right)+R_0+K_2\NoN\\
R_1+R_2&\leq \frac{1}{2}\log\left(1+\INR_1+\frac{\SNR_1}{1+\INR_2}\right)+\frac{1}{2}\log\left(1+\INR_2+\frac{\SNR_2}{1+\INR_1}\right)+2R_0.\NoN\\
2R_1+R_2&\leq \frac{1}{2}\log\left(1+\SNR_1+\INR_1\right)+\frac{1}{2}\log\left(1+\INR_2+\frac{\SNR_2}{1+\INR_1}\right)+\frac{1}{2}\log\left(\frac{1+\SNR_1}{1+\INR_2}\right)+2R_0+K_1\NoN\\
R_1+2R_2&\leq \frac{1}{2}\log\left(1+\SNR_2+\INR_2\right)+\frac{1}{2}\log\left(1+\INR_1+\frac{\SNR_1}{1+\INR_2}\right)+\frac{1}{2}\log\left(\frac{1+\SNR_2}{1+\INR_1}\right)+2R_0+K_2,
\end{align}
where
\begin{align}
K_1&=\frac{1}{2}\log\left(\frac{1+\SNR_1+\SNR_{r1}}{1+\SNR_1}\right)\\
K_2&=\frac{1}{2}\log\left(\frac{1+\SNR_2+\SNR_{r2}}{1+\SNR_2}\right).
\end{align}
\end{theorem}
\begin{corollary}
When $\INR_{i},\SNR_{ri}<\SNR_i,i=1,2$, the capacity region is contained in the region of $(R_1,R_2)$ rate pairs defined by \eqref{eq:bounduseful}.
\end{corollary}

\begin{proof}
The corollary follows from the main theorem, since when $\SNR_{ri}<\SNR_i$, we have $K_i<0.5$, for $i=1,2$. The main theorem can be proved in following steps:
\BEN
\item The first bound is obtained by providing $X_2$ and $Y_r$ to destination 1. From Fano's inequality, we have:
\begin{align}
nR_1&\leq I(X_1^n;Y_1^nX_r^n)+n\epsilon\NoN\\
&\leq I(X_1^n;Y_1^nY_r^n)+n\epsilon\NoN\\
&\leq I(X_1^n;Y_1^nY_r^nX_2^n)+n\epsilon\NoN\\
&=I(X_1^n;Y_1^nY_r^n\big\vert X_2^n)+n\epsilon\NoN\\
&=I(X_1^n;h_{11}X_1^n+Z_1^n,g_1X_1^n+Z_r^n)+n\epsilon\NoN\\
&\leq nI(X_{1g};h_{11}X_{1}^G+Z_1,g_1X_{1}^G+Z_r)+n\epsilon\NoN\\
&=\frac{n}{2}\log(1+\SNR_1)+nI(X_{1}^G;g_1X_{1}^G+Z_r\big\vert h_{11}X_{1}^G+Z_1)+n\epsilon\NoN\\
&=\frac{n}{2}\log(1+\SNR_1)+\frac{n}{2}\log\left(\frac{N+h_{11}^2P_1+g_1^2P_1}{h_{11}^2P_1+N}\right)+n\epsilon\NoN\\
&=\frac{n}{2}\log(1+\SNR_1)+nK_1+n\epsilon\NoN,
\end{align}
where  $X_{1}^G$ representing a Gaussian random variable with variance $P_1$.

\item The second bound can be found by symmetry from the first bound.

\item The third bound is obtained by providing $X_2$ and $Y_r$ to destination 1. Starting from Fano's inequality, we have:
\begin{align}
n(R_1+R_2)&\leq I(X_1^n;Y_1^n X_r^n)+I(X_2^n;Y_2^nX_r^n)+n\epsilon\NoN\\
&\leq I(X_1^n;Y_1^nX_r^nX_2^nY_r^n)+I(X_2^n;Y_2^n)+I(X_2^n;X_r^n|Y_2^n)+n\epsilon\NoN\\
&\leq I(X_1^n;Y_1^nY_r^nX_2^n)+I(X_2^n;Y_2^n)+nR_0+n\epsilon\NoN\\
&= I(X_1^n;Y_1^nY_r^n|X_2^n)+I(X_2^n;Y_2^n)+nR_0+n\epsilon\NoN\\
&=I(X_1^n;Y_1^n|X_2^n)+I(X_2^n;Y_2^n)+I(X_1^n;Y_r^n|X_2^n,Y_1^n)+nR_0+n\epsilon\NoN\\
&\overset{(a)}\leq \frac{n}{2}\log\left(1+\SNR_1\right)+\frac{n}{2}\log\left(1+\frac{\SNR_2}{1+\INR_2}\right)+h\bigl(g_1X_1^n+Z_r^n\big \vert h_{11}X_1^n+Z_1^n\bigr)\NoN\\
&\hspace{8cm} -h(Z_r^n) +nR_0+n\epsilon\NoN\\
&\overset{(b)}\leq \frac{n}{2}\log\left(1+\SNR_1\right)+\frac{n}{2}\log\left(1+\frac{\SNR_2}{1+\INR_2}\right)+nh\bigl(g_1X_{1g}+Z_r\big\vert h_{11}X_{1g}+Z_1\bigr)\NoN\\
&\hspace{8cm}-nh(Z_r)+nR_0+n\epsilon\NoN\\
&=\frac{n}{2}\log(1+\SNR_1)+\frac{n}{2}\log\left(1+\frac{\SNR_2}{1+\INR_2}\right)+K_1+nR_0\NoN
\end{align}
where (a) follows from the so-called Z-channel upper bound of \cite{kramer_outerbounds} and \cite[Equation (45)]{etkin_tse}, and (b) follows from \cite[Lemma 1]{veeravalli-outerbound}.

\item The fourth bound is obtained from the third bound by symmetry.

\item The fifth bound is a trivial extension of the genie-aided bound in \cite[Section 3.4]{etkin_tse} with addition of a relay. Starting from Fano's inequality, we have:
\begin{align}
n(R_1+R_2)&\leq I(X_1^n;Y_1^n X_r^n)+I(X_2^n;Y_2^nX_r^n)+n\epsilon\NoN\\
&\leq
I(X_1^n;Y_1^n)+I(X_2^n;Y_2^n)+2h(X_r^n)+n\epsilon\NoN\\
&\leq I(X_1^n;Y_1^n)+I(X_2^n;Y_2^n)+2nR_0+n\epsilon\NoN\\
&\overset{(a)}\leq
\frac{1}{2}\log\left(1+\INR_1+\frac{\SNR_1}{1+\INR_2}\right)+\frac{1}{2}\log\left(1+\INR_2+\frac{\SNR_2}{1+\INR_1}\right)+2nR_0+n\epsilon.\NoN
\end{align}
where (a) follows from \cite[Section 3.4]{etkin_tse}.

\item The sixth bound is found using the bound on $2R_1+R_2$ \cite[Theorem 3]{etkin_tse}, and providing $X_2,Y_r$ to user one. Starting from Fano's inequality, we have:
\begin{align}
n(2R_1+R_2)&\leq I(X_1^n;Y_1^n X_r^n)+I(X_2^n;Y_2^nX_r^n)+I(X_1^n;Y_1^nX_r^n)+n\epsilon\NoN\\
&\leq
I(X_1^n;Y_1^n)+I(X_2^n;Y_2^n)+2h(X_r^n)+I(X_1^n;Y_1^nY_r^n|X_2^n) +n\epsilon\NoN\\
&\leq I(X_1^n;Y_1^n)+I(X_2^n;Y_2^n)+I(X_1^n;Y_1^nY_r^n|X_2^n)+I(X_1^n;Y_r^n|X_2^nY_1^n)+2nR_0+n\epsilon\NoN\\
&\overset{(a)}
\leq \frac{1}{2}\log\left(1+\SNR_1+\INR_1\right)+\frac{1}{2}\log\left(1+\INR_2+\frac{\SNR_2}{1+\INR_1}\right)+\frac{1}{2}\log\left(\frac{1+\SNR_1}{1+\INR_2}\right)\NoN\\
&\hspace{3cm} + h\bigl(g_1X_1^n+Z_r^n\big \vert h_{11}X_1^n+Z_1^n\bigr) -h(Z_r^n) +2nR_0+n\epsilon\NoN\NoN\\
&\overset{(b)}\leq \frac{1}{2}\log\left(1+\SNR_1+\INR_1\right)+\frac{1}{2}\log\left(1+\INR_2+\frac{\SNR_2}{1+\INR_1}\right)+\frac{1}{2}\log\left(\frac{1+\SNR_1}{1+\INR_2}\right)\NoN\\
&\hspace{3cm} + nh\bigl(g_1X_{1g}+Z_r\big\vert h_{11}X_{1g}+Z_1\bigr)-nh(Z_1) +2nR_0+n\epsilon\NoN\\
&= \frac{1}{2}\log\left(1+\SNR_1+\INR_1\right)+\frac{1}{2}\log\left(1+\INR_2+\frac{\SNR_2}{1+\INR_1}\right)+\frac{1}{2}\log\left(\frac{1+\SNR_1}{1+\INR_2}\right)\NoN\\
&\hspace{8cm} + nK_1 +2nR_0+n\epsilon\NoN.
\end{align}
where (a) follows from the bound on $2R_1+R_2$ in \cite[Theorem 3]{etkin_tse}, and (b) follows from \cite[Lemma~1]{veeravalli-outerbound}.
\item The bound on $R_1+2R_2$ is obtained from the bound on $R_1+2R_2$ by switching the 1 and 2 indices.
\EEN
\end{proof}


\section{Asymptotic Analysis of Symmetric Sum Rate\label{sec:CFanalize}}
In this appendix, we investigate how fast the sum rate improvement using GHF and CF scales with respect to the capacity of a point-to-point Gaussian channel $0.5\log\SNR$, using a relay link of rate $R_0$ as $\SNR$ grows.  To this end, we let $R_0=0.5\cdot\rho\cdot\log\SNR$ and let $\SNR\ra\infty$ as $N\ra 0$ for fixed $h_{11},h_{22},g_1,g_2$ while $\alpha_1,\alpha_2$ are fixed. This asymptotic scenario corresponds to an interference channel where $\abs{h_{ii}},\abs{g_i}\gg \abs{h_{12}},\abs{h_{21}}\gg N$ for $i=1,2$. Since $g_1,g_2$ are fixed and nonzero, we  find that asymptotically  as $N\ra 0$, $\beta_1,\beta_2\ra 1^-$. To  simplify the problem, we also focus on the symmetric case where $\alpha_1=\alpha_2=\alpha$.
\label{sec:app:cf}

\subsection{GHF\label{sec:cf0}}
Note that although Theorem~\ref{thm:improvement}
characterizes the capacity region to within a constant gap under some constraints on $R_0$, we can still use the relay strategy of Theorem~\ref{thm:improvement} to obtain an achievable rate for larger $R_0$ beyond the constraints in \eqref{eq:r0cond}. To find the asymptotic achievable sum rate, from (\ref{eq:d-e-limits}) and for $\alpha_1=\alpha_2=\alpha$,  we have:
\begin{subequations}\label{eq:a-e-asym}
\begin{align}
a_1&\ra\frac{1}{2}(1-\alpha_2)\log\SNR,\\
\intertext{and}
g_1&\ra \frac{1}{2}\log\left(\SNR\right),\\
\intertext{and}
e_1&\ra\frac{1}{2}\max\{\alpha_1,1-\alpha_2\}\log\SNR
\end{align}
\end{subequations}
Switching indices, we also obtain asymptotic first-order expansions for $a_2,g_2,e_2$.

Next, we can also compute $\Delta a_i, \Delta g_i, \Delta e_i$ asymptotically as $\SNR\ra \infty$ for fixed $\rho,\alpha_i,\beta_i,i=1,2$. We have
\begin{align}\label{eq:g1asym}
\Delta{g}_1&=\min\bigl\{R_0,I(Y_r;\hat{Y}_r|Y_1)\bigr\},\NoN\\
&=\min\bigl\{0.5\rho\log\SNR,I(Y_r;\hat{Y}_r|Y_1)\bigr\}\NoN
\intertext{and}
&I(Y_r;\hat{Y}_r|Y_1)\NoN\\
&= \frac{1}{2} \log \left( 1 + \frac{
	(g_1h_{22}-g_2h_{12})^2P_1P_2 + c_2 N}
	{(h_{12}^2P_1+h_{22}^2P_2+N) q} \right)\NoN\\
&\ra \frac{1}{2}\log\left(1+\theta_1\frac{N\cdot\SNR^2}{q(\SNR+\INR_1)}\right)\NoN\\
&\ra \frac{1}{2}\log\left(1+\theta_1\frac{N\cdot\SNR}{q}\right)\NoN\\
&\overset{(a)}\ra \frac{1}{2}\log\left(1+\frac{N\cdot\SNR}{q}\right)
\end{align}
where (a) follows since $\theta_1$ can be found asymptotically to be a constant:
\begin{align}
\theta_1&=\norm{\frac{g_1h_{21}-g_2h_{11}}{h_{11}h_{22}}}\NoN\\
&\ra \norm{\frac{g_2}{h_{22}}}.
\end{align}
A similar asymptotic expression can be found for $\Delta g_2$ by switching indexes 1 and 2.

Next, following derivations similar to \eqref{eq:compe1}, we can also find asymptotic first-order expressions for $\Delta e_1$ and $\Delta e_2$. We have:
\begin{align}\label{eq:e1asym}
\Delta e_1&=\min\{R_0,I(\hat{Y}_r;Y_r\vert Y_1W_1)\}\NoN\\
&=\min\{0.5\rho\log\SNR,I(\hat{Y}_r;Y_r\vert Y_1W_1)\}\NoN\\
\intertext{and,}
&I(\hat{Y}_r;Y_r|Y_1W_1)\NoN\\
&\ra\frac{1}{2}\log\left(1+\norm{\frac{g_1h_{21}-g_2h_{11}}{h_{11}h_{22}}}\cdot\frac{N\cdot\SNR^2}{q\bigl(\SNR+\INR_1\cdot\INR_2+\INR_2\bigr)}\right)\NoN\\
&=\frac{1}{2}\log\left(1+\theta_1\frac{N\cdot\SNR^2}{q\bigl(\SNR+\SNR^{\alpha_1+\alpha_2}+\SNR^{\alpha_2}\bigr)}\right)\NoN\\
&\ra\frac{1}{2}\log\left(1+\frac{N\cdot\SNR^2}{q\cdot\max\bigl\{\SNR,\SNR^{\alpha_1+\alpha_2},\SNR^{\alpha_2}\bigr\}}\right)\NoN\\
&\ra\frac{1}{2}\log\left(1+\frac{N\cdot\SNR^2}{q\cdot\max\bigl\{\SNR,\SNR^{\alpha_1+\alpha_2}\bigr\}}\right)\NoN\\
&=\frac{1}{2}\log\left(1+\frac{N\cdot\SNR^{2-\max\{1,\alpha_1+\alpha_2\}}}{q}\right)
\end{align}
For $\Delta a_1$, we have:
\begin{align}\label{eq:a1asym}
{\Delta}a_1=&\min\{R_0,I(\hat{Y}_r;Y_r\vert W_1Y_1W_2)\}\NoN\\
\intertext{and}
&I(\hat{Y}_r;Y_r\vert W_1Y_1W_2)\NoN\\
&=I(Y_r+\eta;Y_r \vert W_1Y_1W_2)\NoN\\
&=I(g_1V_1+g_2V_2+Z_r;g_1V_1+g_2V_2+Z_r+\eta \vert h_{11}V_1+h_{21}V_2+Z_1)\NoN\\
&=\frac{1}{2}\log\left(1+\frac{N+\var\bigl(g_1V_1+g_2V_2\vert h_{11}V_1+h_{21}V_2+Z_1\bigl)}{q}\right)\NoN\\
&=\frac{1}{2}\log\left(1+\frac{N}{q}+\frac{\norm{g_1h_{21}-g_2h_{11}}P_{v1}P_{v2}+N(\norm{g_1}P_{v1}+\norm{g_2}P_{v2})}{\norm{h_{11}}P_{v1}+\norm{h_{21}}P_{v2}+N}\cdot\frac{1}{q}\right)\NoN\\
&\ra\frac{1}{2}\log\left(1+\frac{N}{q}+\frac{N\cdot\cdot\SNR^2/(\INR_1\cdot\INR_2)+N\cdot\SNR_{r1}/\INR_2+N\cdot\SNR_{r2}/\INR_1}{N\cdot\SNR/\INR_2+2N}\cdot\frac{1}{q}\right)\NoN\\
&=\frac{1}{2}\log\left(1+\frac{N}{q}+\frac{\cdot\SNR^2+\SNR_{r1}\cdot\INR_1+\SNR_{r2}\cdot\INR_2}{\SNR\cdot\INR_1+2\INR_1\cdot\INR_2}\cdot\frac{N}{q}\right)\NoN\\
&=\frac{1}{2}\log\left(1+\frac{N}{q}+\frac{\SNR^2}{\SNR^{1+\alpha_1}}\cdot\frac{N}{q}\right)\NoN\\
&\ra\frac{1}{2}\log\left(1+\frac{N}{q}+\frac{N\cdot\SNR^{1-\alpha_1}}{q}\right).
\end{align}
Finally, asymptotically $\Delta_1$ tends to:
\begin{align}
{\Delta}_1&=I(\hat{Y}_r;Y_r\vert X_1Y_1W_2)\NoN\\
&=\frac{1}{2}\log\left(1+\frac{N}{q}+\frac{\norm{g_2}}{q}\cdot\frac{P_{v2}}{2}\right)\NoN\\
&\ra\frac{1}{2}\log\left(1+\frac{N}{q}+\frac{\norm{g_2}P_2}{2q\cdot\INR_1}\right)\NoN\\
&\ra\frac{1}{2}\log\left(1+\frac{N}{q}+\frac{N\cdot\SNR_{r1}}{2q\cdot\INR_1}\right)\NoN\\
&=\frac{1}{2}\log\left(1+\frac{N}{q}+\frac{N\cdot\SNR^{\beta_1}}{2q\cdot\SNR^{\alpha_1}}\right)\NoN\\
&\ra\frac{1}{2}\log\left(1+\frac{N}{q}+\frac{N\cdot\SNR^{1-\alpha_1}}{q}\right)\label{eq:d1asym}
\end{align}
Note that from \eqref{eq:d1asym} and \eqref{eq:a1asym}, we have $\Delta a_1\approx \Delta_1$ asymptotically, which is expected as $W_1\ra X_1$ as $N\ra 0$. Similar expressions for $\Delta a_2,\Delta_2$ are found by switching 1 and 2 indices.

Now, for $q$ given in \eqref{eq:qweak}, we have:
\begin{align}
q&=\max\{N,\norm{g_1}P_{v1},\norm{g_2}P_{2}\}\NoN\\
&=N\max\{1,\frac{\norm{g_1}P_1}{\INR_1},\frac{\norm{g_2}P_{2}}{\INR_2}\}\NoN\\
&=N\max\{1,\SNR^{\beta_1-\alpha_1},\SNR^{\beta_2-\alpha_2}\}\NoN\\
&\ra N\max\{1,\SNR^{1-\alpha_1},\SNR^{1-\alpha_2}\}
\label{eq:qasymgh}.
\end{align}

Substituting \eqref{eq:qasymgh} in the above asymptotic derivations for $a_i,g_i,e_i,\Delta a_i,\Delta g_i,\Delta e_i,i=1,2$ and using (\ref{eq:rateregion}), we find that the 
asymptotic sum rate for the GHF strategy:
\begin{align}\label{eq:app:ghf}
R_1+R_2&\leq a_1+g_2+\Delta a_1+\Delta g_2-\Delta_1-\Delta_2\NoN\\
&\ra \frac{1}{2}(2-\alpha_2)\log\SNR+\frac{1}{2}\min\{\rho,\alpha_1,\alpha_2\}\log\SNR\\
R_1+R_2&\leq a_2+g_1+\Delta a_2+\Delta g_1-\Delta_1-\Delta_2\NoN\\
&\ra \frac{1}{2}(2-\alpha_1)\log\SNR+\frac{1}{2}\min\{\rho,\alpha_1,\alpha_2\}\log\SNR\\
R_1+R_2&\leq e_1+e_2+\Delta e_1+\Delta e_2-\Delta_1-\Delta_2\\
&\ra \frac{1}{2}\max\{\alpha_1+\alpha_2,2-\alpha_1-\alpha_2\}\log\SNR+\min\{\rho,1+\min\{\alpha_1,\alpha_2\}-\max\{1,\alpha_1+\alpha_2\}\}\NoN\\
&=\max\{\alpha_1+\alpha_2,2-\alpha_1-\alpha_2\}\log\SNR+\min\{\rho,\alpha_1,\alpha_2,1-\alpha_1,1-\alpha_2\}.
\end{align}

\subsection{CF with Decoding the Relay Codeword First\label{sec:cf1}}
In this scheme, since both users uniquely decode the quantized relay codeword, the achievable rate region can be computed by replacing $Y_1$ and $Y_2$ with $(Y_1,\hat{Y}_r)$ and $(Y_2,\hat{Y}_r)$. Thus, $R_1+R_2$ satisfying the following constraints are achievable:
\begin{align}\label{eq:cfsumrate}
R_1+R_2&\leq I(Y_1\hat{Y}_r;X_1|W_1,W_2)+I(Y_2\hat{Y}_r;X_2W_1)\NoN\\
R_1+R_2&\leq I(Y_2\hat{Y}_r;X_2|W_1,W_2)+I(Y_1\hat{Y}_r;X_1W_2)\NoN\\
R_1+R_2&\leq I(Y_1\hat{Y}_r;X_1W_2|W_1)+I(Y_2\hat{Y}_r;X_2W_1|W_2)
\end{align}
For Etkin-Tse-Wang power splitting strategy of \eqref{eq:etkinsplit}, the mutual information terms in the above can be simply computed. We have:
\begin{align}
I(X_1;Y_1\hat{Y}_r|W_1W_2)
&=I(X_1;Y_1|W_1W_2)+I(X_1;\hat{Y}_r|Y_1W_1W_2)\NoN\\
&=I(X_1;Y_1|W_1W_2)+I(Y_r;\hat{Y}_r|Y_1W_1W_2)-I(Y_r;\hat{Y}_r|Y_1X_1W_2)\NoN\\
&:={a}_1+\tilde{\Delta} {a}_1-{\tilde{\Delta}}_1
\end{align}

\begin{align}
I(X_1W_2;Y_1\hat{Y}_r)
&=I(X_1W_2;Y_1)+I(X_1W_2;\hat{Y}_r|Y_1)\NoN\\
&=I(X_1W_2;Y_1)+I(Y_r;\hat{Y}_r|Y_1)-I(Y_r;\hat{Y}_r|Y_1X_1W_2)\NoN\\
&:={g}_1+\tilde{\Delta} {g}_1-{\tilde{\Delta}}_1
\end{align}

\begin{align}
I(X_1W_2;Y_1\hat{Y}_r|W_1)
&=I(X_1W_2;Y_1|W_1)+I(X_1W_2;\hat{Y}_r|W_1Y_1)\NoN\\
&=I(X_1W_2;Y_1|W_1)+I(Y_r;\hat{Y}_r|W_1Y_1)-I(Y_r;\hat{Y}_r|Y_1X_1W_2)\NoN\\
&:={e}_1+\tilde{\Delta} {e}_1-{\tilde{\Delta}}_1.
\end{align}
Switching indices gives the remaining terms in \eqref{eq:cfsumrate}.
Now, it is straightforward to characterize the asymptotic behavior of the achievable sum rate.  To analyze the asymptotic rates, we let $\SNR\ra \infty$, while $\alpha_i,\beta_{i}<1$ are fixed, $i=1,2$. We also choose $\rho$ such that $R_0=0.5\cdot\rho\cdot \log \SNR$ satisfies  \eqref{eq:r0cond}.

First, asymptotic values for $a_i,g_i,e_i$ are calculated in \eqref{eq:a-e-asym}. The remaining terms $\tilde{\Delta} a_i,\tilde{\Delta} g_i, \tilde{\Delta} e_i$ can also be computed following the derivations in \eqref{eq:g1asym}--\eqref{eq:d1asym}.

Now, for $q$ given in \eqref{eq:qcf}, we have
\begin{align}
 q&\ra \frac{1}{2^{2R_0}-1}\max\left\{\frac{(g_1h_{21}-g_2h_{11})^2P_1P_2} {(h_{11}^2P_1+h_{21}^2P_2)},\frac{(g_1h_{22}-g_2h_{12})^2P_1P_2} {(h_{12}^2P_1+h_{22}^2P_2)}\right\}\NoN\\
&\ra 2^{-\rho\log\SNR}\max\left\{\theta_1\frac{N\cdot\SNR^2}{\SNR+\INR_1},
\theta_2\frac{N\cdot\SNR^2}{\SNR+\INR_2}\right\},\NoN\\
\end{align}
asymptotically as $\SNR\ra \infty$.

For the above asymptotic value for $q$ and using \eqref{eq:g1asym}--\eqref{eq:d1asym}, we have the following achievable sum rate from \eqref{eq:cfsumrate}:
\begin{align}\label{eq:app:cf1}
R_1+R_2&\leq a_1+g_2+\tilde{\Delta} a_1+\tilde{\Delta} g_2-\tilde{\Delta}_1-\tilde{\Delta}_2\NoN\\
&\ra \frac{1}{2}(2-\alpha_2)\log\SNR+\frac{1}{2}\rho\log\SNR-\frac{1}{2}(\rho-\alpha_2)^+\log\SNR\NoN\\
&= \frac{1}{2}(2-\alpha_2)\log\SNR+\frac{1}{2}\min\{\rho,\alpha_2\}\log\SNR\\
R_1+R_2&\leq a_2+g_1+\tilde{\Delta} a_2+\tilde{\Delta} g_1-\tilde{\Delta}_1-\tilde{\Delta}_2\NoN\\
&\ra \frac{1}{2}(2-\alpha_1)\log\SNR+\frac{1}{2}\min\{\rho,\alpha_1\}\log\SNR\\
R_1+R_2&\leq e_1+e_2+\tilde{\Delta} e_1+\tilde{\Delta} e_2-\tilde{\Delta}_1-\tilde{\Delta}_2\NoN\\
&\ra \frac{1}{2}\max\{\alpha_1+\alpha_2,2-\alpha_1-\alpha_2\}\log\SNR+\Bigl(\rho+1-\max\{1,\alpha_1+\alpha_2\}\Bigr)^+\log\SNR\NoN\\
&\quad -\frac{1}{2}(\rho-\alpha_1)^+\log\SNR-\frac{1}{2}(\rho-\alpha_2)^+\log\SNR.
\end{align}

\subsection{CF with Decoding the Relay Codeword Second \label{sec:cf2}}
With this strategy, each destination first decodes its own common message, and then uses this message as additional side information to decode the relay observation. Decoding of $\hat{Y}_r$ with this decoding order is successful if \eqref{eq:cf2wynerziv} holds. To satisfy \eqref{eq:cf2wynerziv},
the relay quantizes its observation $Y_r$ using an auxiliary variable $\hat{Y}_r=Y_r+\eta$ with $\eta\sim \N(0,q)$  where  $q$ is given as
\begin{align}
q&=\frac{1}{2^{2R_0}-1}\max\bigl\{\var(Y_r\vert Y_1W_1),\var(Y_r\vert Y_2W_2)\bigr\}\label{eq:qgoodcf}\\
&\ra \frac{1}{2^{2R_0}-1}\max\left(\frac{(g_1h_{21}-g_2h_{11})^2P_{v1}P_2} {(h_{11}^2P_{v1}+h_{21}^2P_2)},\frac{(g_1h_{22}-g_2h_{12})^2P_1P_{v2}} {(h_{12}^2P_1+h_{22}^2P_{v2})}\right)\NoN\\
&\ra\frac{1}{{2^{2R_0}-1}}{\max\left(\theta_1\frac{N\cdot\SNR^2}{\SNR+\INR_1\cdot\INR_2},\theta_2\frac{N\cdot\SNR^2}{\SNR+\INR_1\cdot\INR_2}\right)},\NoN\\
&\ra N\cdot\SNR^{-\rho}\cdot\SNR^{\min\{1,2-2\alpha\}}\label{eq:qcf2asym}
\end{align}

We now compute the asymptotic sum rate for decoding order  $W_1\ra \hat{Y}_r\ra W_2 \ra V_1$ in the symmetric case where $\alpha_1=\alpha_2$. To decode $W_1$ first at destination 1, we need
\begin{align}\label{eq:gcf6}
T_1&< I(W_1;Y_1)\NoN\\
&=I(W_1;h_{11}(W_1+V_1)+h_{21}X_2+Z_1)\NoN\\
&=\frac{1}{2}\log\left(1+\frac{h_{11}^2P_{w1}}{N+\norm{h_{11}}P_{v1}+h_{21}^2P_2}\right)\NoN\\
&=\frac{1}{2}\log\left(1+\frac{\SNR-\SNR/\INR_2}{1+\INR_1+\SNR/\INR_2}\right)\NoN\\
&=\frac{1}{2}\log\left(1+\frac{\SNR-\SNR^{1-\alpha}}{1+\SNR^{\alpha}+\SNR^{1-\alpha}}\right)\NoN\\
&\ra\frac{1}{2}(1-\alpha)^+\log \SNR,
\end{align}
asymptotically as $\SNR\ra \infty$, where $T_1$ denotes the rate of the common message encoded by $W_1$.

With $\hat{Y}_r$ decoded, the decoder first decodes $W_2$ and then decodes the remaining private message. Decoding of $W_2$ is successful provided that:
\begin{align}
T_2&\leq I(W_2;Y_1\hat{Y}_r\big \vert W_1)\NoN\\
&=I(W_2;Y_1|W_1)+I\bigl(\hat{Y}_r;Y_r\big \vert Y_1W_1\bigr)-I\bigl(\hat{Y}_r;Y_r\big \vert Y_1W_1W_2\bigr)\label{eq:sd4cf}
\end{align}
where $T_2$ is the rate of common message encoded by $W_2$. In the asymptotic regime, $I\bigl(\hat{Y}_r;Y_r\big \vert Y_1W_1W_2\bigr)$ is computed by substituting \eqref{eq:qcf2asym} in \eqref{eq:a1asym}, which yields:
\begin{align}
I\bigl(\hat{Y}_r;Y_r\big \vert Y_1W_1W_2\bigr)=\frac{1}{2}\Bigl(\rho-\min\{\alpha,1-\alpha\}\Bigr)^+\log\SNR.
\end{align}
We can similarly find the asymptotic value of $I\bigl(\hat{Y}_r;Y_r\big \vert Y_1W_1\bigr)$ by substituting \eqref{eq:qcf2asym} in \eqref{eq:e1asym}, which gives:
\begin{align}
I\bigl(\hat{Y}_r;Y_r\big \vert Y_1W_1W_2\bigr)=\frac{1}{2}\rho\log\SNR.
\end{align}
Finally, the asymptotic value of the remaining term $I(W_2;Y_1|W_1)$ can be found as:
\begin{align}
I(W_2;Y_1|W_1)&=\frac{1}{2}\log\left(1+\frac{\INR_1-1}{1+\SNR/\INR_2}\right)\NoN\\
&\ra\frac{1}{2}\log\left(1+\frac{\INR_1\cdot\INR_2}{\SNR}\right)\NoN\\
&= \frac{1}{2}(2\alpha-1)^+\log\SNR\label{eq:gcfb2}.
\end{align}

Next, the decoder decodes $V_1$ by subtracting $W_1,W_2$, and without the relay help. The asymptotic rate of private message is given by:
\begin{align}
S_1\leq I(X_1;Y_1|W_1W_2)=a_1&\ra \frac{1}{2}(1-\alpha)\log\SNR. \label{eq:gcfb3}
\end{align}

Thus, we get the following constraints for $T_1,S_1$:\begin{align}
T_1&\leq \frac{1}{2}(1-\alpha)\log \SNR\NoN\\
T_1&\leq \frac{1}{2}(2\alpha-1)^+\log \SNR +\frac{1}{2}\rho\log\SNR- \frac{1}{2}\Bigl(\rho-\min\{\alpha,1-\alpha\}\Bigr)^+\log\SNR\NoN\\
S_1&\leq \frac{1}{2}(1-\alpha)\log \SNR\NoN,
\end{align}
which result in the following asymptotic achievable rate for user one:
\begin{align}
R_1&\leq \frac{1}{2}(2-2\alpha)\log\SNR\NoN\\
R_1&\leq \frac{1}{2}(1-\alpha+(2\alpha-1)^+)\log\SNR+\frac{1}{2}\rho\log\SNR- \frac{1}{2}\Bigl(\rho-\min\{\alpha,1-\alpha\}\Bigr)^+\log\SNR\NoN\\
&= \frac{1}{2}\max\{\alpha,1-\alpha\}\log\SNR+\frac{1}{2}\rho\log\SNR- \frac{1}{2}\Bigl(\rho-\min\{\alpha,1-\alpha\}\Bigr)^+\log\SNR\NoN.
\end{align}
A similar set of constraints are found for $R_2$, and thus, we have the following asymptotic achievable sum rate for CF with modified decoding order:
\begin{align}
R_1+R_2&\leq (2-2\alpha)\log\SNR\NoN\\
R_1+R_2&\leq \max\{\alpha,1-\alpha\}\log\SNR+\rho\log\SNR- \Bigl(\rho-\min\{\alpha,1-\alpha\}\Bigr)^+\log\SNR\NoN.
\end{align}



\bibliographystyle{IEEEtran}
\bibliography{IEEEabrv,reference}

\end{document}